\documentclass{aa}  
\usepackage{graphicx,color}
\usepackage{txfonts}
\usepackage{hyperref}
\bibpunct{(}{)}{;}{a}{}{,} % to follow the A&A style

\newcommand{\msun}{{$M_\odot$}}
\newcommand{\msm}{{M_\odot}}

\newcommand{\kms}{{km~s$^{-1}$}}
\newcommand{\sbr}[1]{\left[ #1 \right]} % square bracket
\newcommand{\feh}[1]{{$\sbr{\mathrm{Fe/H}}=#1$}}

\newcommand{\rb}[1]{\textcolor{black}{ #1 }}
\newcommand{\rvtwo}[1]{#1}
\newcommand{\rv}[1]{#1}

\newcommand{\eql}[2]{\begin{equation} \label{#1} #2 \end{equation}}

\begin{document}

%\title{Can hypernovae or variations of the initial mass function 
%lead to gas expulsion in young globular clusters?}
\title{Gas expulsion in massive star clusters?}
\titlerunning{Gas expulsion in massive star clusters?}
\subtitle{\rv{Constraints from observations of young and gas-free objects}}
 \author{Martin G. H. Krause \inst{1,2,3} \fnmsep\thanks{E-mail:
    krause@mpe.mpg.de} 
  \and Corinne Charbonnel \inst{4,5}
  \and Nate Bastian \inst{6}
  \and Roland Diehl \inst{2,3} 
   %\and Andreas Burkert \inst{3,1}
%\and Ortwin Gerhard \inst{2} \and Jochen Greiner \inst{2}
%\and Karsten Kretschmer \inst{4}
}

\institute{
Universit\"ats-Sternwarte M\"unchen, Ludwig-Maximilians-Universit\"at,
Scheinerstr. 1, 81679 M\"unchen, Germany
\and  Max-Planck-Institut f\"ur extraterrestrische Physik,
  Giessenbachstr.~1, 85741 Garching, Germany 
\and Excellence Cluster Universe, Technische Universit\"at
  M\"unchen, Boltzmannstrasse 2, 85748 Garching, Germany 
\and Geneva Observatory, University of Geneva, 
     51 Chemin des Maillettes, 1290 Versoix, Switzerland
\and  IRAP, UMR 5277 CNRS and Universit\'e de Toulouse, 14 Av. E. Belin, 31400 Toulouse, France
\and Astrophysics Research Institute, Liverpool John Moores University, 146 Brownlow Hill, Liverpool L3 5RF, UK
%\and Universit\"atssternwarte M\"unchen, Scheinerstr.~1, 81679 M\"unchen 
%\and Fran\c{c}ois Arago Centre, APC, Universit\'e Paris Diderot, F-75013 Paris, France 
}

   \date{Received June 8, 2015; accepted ?}

% \abstract{}{}{}{}{} 
% 5 {} token are mandatory
 
  \abstract
   {Gas expulsion is a central concept in \rb{some of the} models for 
	\rv{multiple populations} and the 
	light-element anticorrelations in globular clusters. 
	\rb{If the star formation efficiency was around 30 per cent and the 
	gas expulsion happened on the crossing timescale, this process could} 
	expel preferentially stars \rv{born with the chemical composition 
	of the proto-cluster gas, while stars with special composition 
	born in the centre would remain bound.} 
	\rv{Recently, a sample of extragalactic, gas-free, young massive clusters has been identified
	that has the potential to test the conditions for gas expulsion.}}
   {We investigate \rv{the conditions required for residual gas expulsion 
	on the crossing timescale. 
	\rv{We consider a standard initial mass function and} different models for the 
	energy production in the cluster:} metallicity dependent 
	stellar winds, radiation, supernovae 
	\rv{and more energetic events}, namely
	winds of young pulsars and the gamma ray burst related hypernovae.
	The latter \rv{may be more energetic than supernovae}
	by up to two orders of magnitude.}
   {We compute a large number of thin shell models for the gas dynamics, 
	and calculate if the Rayleigh-Taylor instability is able
	to disrupt the shell before it reaches \rv{the} escape speed.
%	\rv{We adress a specific sample of young massive clusters, and compute 
%	a model grid.}
}
   {
\rv{We show that the success of gas expulsion depends on the compactness index of a star cluster 
	$C_5 \equiv  (M_*/10^5 M_\odot)/(r_\mathrm{h}/\mathrm{pc})$, 
	with initial stellar mass $M_*$ and half-mass radius $r_\mathrm{h}$.
	For given $C_5$,
	a certain critical, local star formation efficiency is required to remove the 
	rest of the gas. Common stellar feedback processes may not lead to gas expulsion
	with significant loss of stars above $C_5 \approx 1$. Considering pulsar winds and hypernovae,
	the limit increases to $C_5 \approx 30$.
	If successful, gas expulsion generally takes place on the 
	crossing timescale.\\
	Some observed young massive clusters have $1<C_5<10$ and are gas-free at
	$\approx 10$~Myr. This suggests that gas expulsion does not affect their stellar mass
	significantly, unless powerful pulsar winds and hypernovae are common in such objects.\\
	By comparison to observations, we show that $C_5$ is a better predictor for the
	expression of multiple populations than stellar mass. The best separation between
	star clusters with and without multiple populations is achieved by a stellar winds-based
	gas expulsion model, where gas expulsion would occur exclusively in star clusters
	without multiple populations. Single and multiple population clusters also have 
	little overlap in metallicity and age. 	}
}
   {\rv{Globular clusters should initially have $C_5 \lesssim 100$, if the gas expulsion 
	paradigm was correct. Early gas expulsion, which is suggested 
	by the young massive cluster observations, hence would require special 
	circumstances, and is excluded for several objects. Most likely,
	the stellar masses did not change significantly at the removal of the primordial gas.
	Instead, the predictive power of the $C_5$ index for the expression 
	of multiple populations is consistent with the idea that gas expulsion  
	may prevent the expression of multiple populations. 
	On this basis, compact young massive clusters should also have multiple populations.} }

   \maketitle

\defcitealias{Bastea14b}{BHC14}
\defcitealias{Marksea12}{M12}
\defcitealias{Krausea12a}{Paper~I}

%%%%%%%%%%%%%%%%%%%%%%%%%%%%%%%%%%%%%%%%%%%%%%%%%%%%%%%%%%%%%%
\section{Introduction}\label{sec:intro}
%%%%%%%%%%%%%%%%%%%%%%%%%%%%%%%%%%%%%%%%%%%%%%%%%%%%%%%%%%%%%%
The formation of stars is associated with radiation, stellar winds,
and energetic explosions. This feedback should usually clear away surrounding
gas at ease \citep[e.g.][]{FHY03,Acrea12,Fierlea12a,Krausea13a,Krumholz14a}.
Indeed, star-forming regions are observed to be actively clearing away gas at an age of
about $10^6$ years \citep[e.g.][]{Vossea10,Vossea12,Preibea12,Galvea13,Klaassea15}, 
and to be free of dense gas from about a few Myr   
after the onset of star formation \citep[e.g., the review by][]{LL03}.   

The gravitational binding energy of a star-forming region increases, however, with the 
square of the total mass, whereas feedback processes are directly proportional to the latter.
One expects therefore a certain mass limit, above which the released energy
is insufficient to remove the remaining gas
\rv{\citep[compare, e.g., ][]{Decrea10,Krausea12a}}. Candidate regions for this to occur
certainly include nuclear star clusters in galaxies, where typically a 
super-massive black hole adds to the gravity of the stars
\citep[e.g.][]{Schartea10}, but also massive star clusters, especially globular clusters (GCs).

If at the time of gas clearance the gravitational potential of such clusters is dominated 
by this gas\rv{, and if the gas clearance happens sufficiently fast, comparable to the crossing timescale}, 
then many stars if not all (disruption) are lost
as well due to the drastic change in gravitational potential \citep[e.g.,][]{BG06,GB06,Bastea08,GiBa08,BKP08,Decrea10,Marksea12,PfaKa13,Pfalzea14}. 
This is commonly referred to as {\em gas expulsion}
\rv{\citep[for a review]{PZMcMG10}}.

%The concept has at one point
%received much attention for all star clusters, because the number
%of embedded (forming) clusters in the Milky Way significantly exceeds the number of 
%older open clusters \citep{LL03}. 
%%The latter, however, appears not be as strong a conclusion as at first thought, 
%%due to the differing definitions of "cluster" when it is embedded or when 
%%it is an exposed open cluster - e.g., \citet{Bressea10}.
%\rv{This is}, however, questioned in the recent literature 
%\rv{\citep[e.g.][]{Kruijea12c}},
%and tidal interactions with molecular clouds have been suggested as alternative mechanism 
%to shape star clusters .

Gas expulsion is central in explanations for the chemical peculiarities in GCs,
and is invoked in certain scenarios as a possibility to eject \rv{a large number of less tightly 
bound first population stars born with the pristine composition of the proto-cluster gas. At the opposite, second population stars showing peculiar abundance properties (e.g. the O-Na anticorrelation) and born in the cluster centre would remain bound after gas expulsion
\citep{DErcolea08,DErcolea10,Krausea13b}. This is one possibility to solve the so-called mass-budget problem to explain the number distribution of stars along the O-Na anticorrelation. Other possibilities exist though, that call for \rvtwo{a} modified \rvtwo{initial mass function (IMF)} 
\citep{PC06,Charbea14}.}
The stars ejected from GCs might form a substantial population of galactic halos
\citep[e.g.,][]{SC11,Larsenea14b}.

Young massive clusters (YMCs) are important test cases for the gas expulsion idea,
because the involved timescales, masses and energetics may be constrained 
\rv{directly by} observations \citep[e.g.,][]{PZMcMG10}.
\rv{Some extragalactic objects appear to be gas-free after 10~Myr or less \citep{Bastea14b,Hollyhea15} 
and are thus very interesting
objects for comparison to theoretical expectations.}

We have investigated gas expulsion in the context of models that aim to explain 
the light-element anticorrelations in GCs with massive progenitors
\citep[Paper~I]{Krausea12a}. We found that for common cluster parameters,
gas expulsion would not work, because the outflow would be disrupted by instabilities
before escape speed was reached. An unusually strong power source,
like the coherent onset of Eddington-strength accretion on all neutron stars and stellar black holes in a cluster would be required to still allow for gas expulsion.
However in that case, gas expulsion would occur relatively late in the cluster life, at the end of the turbulent SNe phase (i.e., typically at ages ~ 40 Myrs), which is incompatible with the fact that YMCs appear to be gas-free after $\approx$4~Myrs.

Here, we put our modelling in a more general context. We first review constraints
on gas expulsion from observations of stellar kinematics, 
light-element anticorrelations, the population of halo stars, and
gas (or the absence thereof) in YMCs  in Sect.~\ref{sec:obsge}.
We describe our methodology in Sect.~\ref{s:method}), and present 
the results in Sect.~\ref{s:results}.
We present detailed gas expulsion models for a sample of YMCs from 
\citet{Bastea14b} in Sect.~\ref{s:YMCs}. 
%\ra{An energy source that has been ignored so far is related to rapid
%spin-down of dark remnants
%(neutron stars and black holes), which manifests observationally in pulsar winds
%\citep[e.g.,][]{Neronea12} and long gamma ray bursts (GRBs) with associated
%type~Ic supernovae \citep[also hypernovae, e.g.,][]{Cano13}.}
%%Type~Ic supernovae associated with Gamma ray bursts (GRBs) may reach kinetic energies beyond 
%%$10^{52}$~erg \citep[hypernovae, e.g.,][]{Cano13,Mazea14,Heesea15}, 
%%much more than standard supernovae.
%These objects are thought to be associated with fast rotating massive stars 
%\citep[e.g.][]{MM12}, and 
%therefore might be expected to occur frequently in young GCs, if the FRMS scenario was correct.
In Sect.~\ref{s:grid}, we show a grid of calculations for various feedback scenarios. 
 We discuss the results in Sect.~\ref{s:disc} and summarise the conclusions in
Sect.~\ref{s:conc}. 
%\rv{All our calculations are performed assuming a Salpeter initial mass function.
%We defer deviations from this assumption to future work.}

\section{Constraints on gas expulsion}\label{sec:obsge}

\subsection{Star cluster \rv{mass functions and} kinematics}

\rv{\citet{LL03} found that "the embedded cluster birthrate exceeds that of visible open clusters by an order of magnitude or more". They concluded
that many clusters were dissolved at an early age, consistent with gas expulsion
ideas. \citet{Bressea10} then determined the distribution of the densities of young
stellar objects in the solar neighbourhood, and found a continuous distribution
around 22 stars pc$^{-2}$. They concluded that the number of embedded
star clusters strongly depends on the adopted star density threshold, complicating
conclusions about early star cluster dissolution. They also found, however,
that class~I (younger) objects have, on average, larger densities than class~II (older)
objects, supporting the idea that star-forming regions often dissolve.
The environment has a strong effect on the evolution of the cluster mass functions 
\citep[see the review by][]{Longmea14}. It is therefore difficult to constrain gas expulsion
in this way.}

\rv{Young ($\approx 10$ Myr), exposed star clusters often appear to have 
supervirial velocity dispersions \citep{Gielea10}. 
This has been discussed as evidence for dissolution after gas expulsion
\citep[e.g.,][]{GB06}. It is, however, expected that many clusters
would have re-virialized by the time of observation \citep{BK07,Gielea10,PZMcMG10}.
\rvtwo{A}n interpretation in terms of a large contribution
from binaries to the velocity dispersion
\citep[compare, e.g., ][]{Leighea15,OKP15} seems more probable
\citep{Gielea10,Rochea10,Cottea12,Clarksea12,HenBruea12,CoHe14}.
Therefore, it is also difficult to constrain gas expulsion in this way.
}

\subsection{Light-element anticorrelations in globular clusters}\label{sec:legc}
With the former tailwind from the general research on star clusters, 
gas expulsion assumed a central role for the understanding  
of GCs \citep[e.g.,][]{Goodwin97a,Goodwin97b,Fennea04,PC06,ParmenGil07,
BKP08,DErcolea08,Decrea10,MK10,Vespea10,SC11,Marksea12}.
In contrast to the vast majority of open (lower mass)
clusters studied so far, \rb{almost all of the} GCs show chemical peculiarities,
most prominently the light-element anti-correlations 
\citep[][for reviews]{Charb10,GCB12}, which 
%for 
%a standard initial mass function (IMF) and a few other reasonable assumptions 
%imply much larger initial cluster masses.
%Gas expulsion might be a natural explanation for this 
%{\em mass budget problem} and the appearance of the 
%light-element anti-correlations 
may well point to fundamental
differences in the mode of star formation.

The light-element anti-correlations in GC stars, most prominently between
Na and O,
together with the constancy of other elements like, e.g., Fe imply
that hydrogen-burning products from more massive stars than the currently 
present ones have been reprocessed, possibly mixed with remaining gas from the
star formation event to form the presently observed low-mass stars
\citep{Grattea01,GCB12,PC06,Prantzea07,Charb10}.
\rvtwo{O-rich and Na-poor stars form the first population,
whereas the O-poor and Na-rich stars constitute the second one.}
Intermediate-mass Asymptotic Giant Branch (AGB) stars
\citep[$\approx 6$~\msun - 11~\msun][]{Ventea01,Ventea13,VenturaDAntona09,VenturaDAntona11, DErcolea10},
and fast rotating massive stars (FRMS)
\rv{\citep[$\gtrsim 25$~\msun][]{MM06,PC06,Decrea07a,Krausea13b,Charbea14}},
variants and combinations of these object classes
\citep{deMinkea09,SG10,Bastea13a,CS14}, and recently also supermassive stars
\citep{DH14} have received attention as possible sources
of the enriched material.
From the nucleosynthesis point of few, FRMS, 
or massive stars in binary systems \citep{deMinkea09}, 
are perhaps the most straightforward
polluter candidates: they produce the Na-O anticorrelation directly
\citep{Decrea07a}, as opposed to AGB stars, where the direct correlation in their 
ejecta has to be turned into an anticorrelation by a precisely  prescribed mixing procedure
\citep{Ventea13}. 
%The FRMS scenario would not need any dilution to explain the Na-O anticorrelation
%\citep{Charbea14}. The observations of the light elements Li and Be in the long-lived low-mass stars \citep[e.g.][]{Pasquea05,Pasquea07,Dobrea14} require however that 
%pristine gas (i.e. unprocessed by massive stars) plays some role in the formation of the
%enriched, second generation stars.

Because the ejecta mass is small in the AGB as well as the FRMS scenario,
gas expulsion has been identified as a possibility to enhance the ratio
of second to first generation stars
\citep[mass budget problem,][]{DErcolea08,DErcolea11,CabZiea15}:
If the second generation was formed close to the massive stars in a mass-segregated
star cluster, the less tightly bound first generation would have been lost preferentially
\citep{Decrea10}.
%Gas expulsion would occur before the formation of the second generation 
%stars in the AGB scenario, and therefore would 
%have the additional advantage of removing the 
%Fe-rich SN~II ejecta, such that the later formed second generation stars 
%would still have the same Fe abundance, as observed in most GCs.

\rv{For these models to work, i.e. remove $\gtrsim 95$~per cent of the stars,
gas expulsion needs to be explosive \citep{Decrea10,KhalBaum15}, i.e.
happen on the crossing timescale of the cluster. According to our models
\citep{Krausea12a,Krausea13b}, stellar winds and supernovae do not
provide enough power for this process. Energy release from the
combined accretion on to the
dark remnants
of the massive stars (neutron stars and black holes) might perhaps
accomplish the task.
This would then, however, happen late, about 35 Myr after cluster formation, when the massive stars would have 
turned into such dark remnants.
}

\rv{If gas expulsion would work in multiple population clusters, one would expect 
a correlation between the total stellar mass and the fraction of enriched stars,
which is, however, not seen \citep{KhalBaum15,BL15}.}

%we have investigated the stability
%of such a scenario against Rayleigh-Taylor modes. We found that this is a particular problem
%for star clusters, because in order to overcome the self-gravity of stars and gas, the shell
%must first be pushed comparatively hard. When the half-mass radius is overcome,
%the shell inevitably accelerates due to the diminishing gravitational pull. Acceleration
%then causes large Rayleigh-Taylor modes to grow. For a generic star cluster with
%$9\times 10^6$~\msun initial gas mass, we found that feedback scenarios based 
%stellar winds and supernovae lead to destruction of the supershell before the escape
%velocity could be reached, and hence gas expulsion would not work. The only way we found
%to expell the gas successfully was via a sudden onset of strong energy injection, which could
%be provided by energy release due to accretion on all the dark remnants, probably
%a quite extreme assumption.

%\subsection{Observational constraints on gas expulsion}
%
%The loss of stars  as well as the presence of gas can be constrained by direct observations 
%of extragalactic globular clusters (XGCs) and young massive clusters (YMCs), 
%if they are indeed the same kind of object as GCs.

\subsection{Extragalactic globular clusters and stars in galactic haloes}
\rv{Extragalactic globular clusters (XGCs)} are very similar to GCs in age and metallicity \citep[for a review]{BS06},
though there are also some differences \citep[e.g.,][]{Montea14}. 
Associations of XGCs with tidal streams suggest that the majority of globular clusters 
was formed in dwarf galaxies and accreted with them to the haloes of bigger galaxies
\citep{Mackea10,Mackea13,Mackea14,Huxea14,Veljea14}.

\rv{Old GCs in the Large Magellanic Cloud have been shown to exhibit the 
O-Na anticorrelation \citep{Mucciea09a}.}
From integrated spectra of nearby dwarf galaxies,
\citet{Larsenea14a,Larsenea14b} find evidence for Na spreads,
consistent with the presence of the Na-O anticorrelation and other GC characteristics.  
Comparing the masses
of XGCs to the total mass of halo stars with the same metallicity in the respective host
galaxies, the latter authors constrain the ejection of first generation low-mass stars to four times 
the current cluster mass, a much tighter constraint than from
the analogous argument for the Milky Way \citep{SC11}. 
This number has to accommodate secular effects as well as 
star clusters that have been disrupted completely. 
Unless stars ejected from the GCs in Fornax had also enough kinetic energy to disappear
from the parent dark-matter halo, this puts tight constraints on gas expulsion.

\subsection{Young massive star clusters}
YMCs \citep[for reviews]{PZMcMG10,Longmea14} provide quite possibly very interesting 
complementary insight into what happened in the infancy of GCs. They reach masses up to 
$10^8$~\msun, which is comparable to the initial mass postulated for GCs from
self-enrichment considerations, and have core radii of one to a few parsec, similar to GCs.
A possible difference concerns the metallicity, which reaches down to lower 
values in the case
of GCs. Some GCs, however, show the anticorrelation, but have higher metallicities
than some YMCs. 
Metallicity affects the thermodynamics of the gas, which may have a rather strong effect 
on the star formation process \citep[e.g.,][]{Petersea12}. It has not yet been possible 
to determine, if an 
Na-O anticorrelation such as characteristic for GCs is present in YMCs.

\rv{Star formation and gas has been searched for in YMCs by
multiwavelength studies including 
H$\alpha$ and infrared \citep{Whitmea11,Bastea14b,Hollyhea15}.
Embedded clusters are routinely detected.
\citet{Whitmea11} and \citet{Hollyhea15} demonstrate for a large
sample that in massive clusters up to $\approx 10^5$~\msun 
there is no gas present after about 4~Myr. 
\citet{Bastea14b} show for a small sample around $10^6 \msm$
and ages from 4 to 15~Myr that gas is not present, entirely consistent
with the studies at lower masses.
Thus, around 4~Myr the gas has either been expelled, shed
more gently in a mass-loaded wind, or it has been very 
efficiently transformed into stars.}

\section{Methods}\label{s:method}
%%%%%%%%%%%%%%%%%%%%%%%%%%%%%%%%%%%%%%%%%%%%%%%%%%%%%%%%%%%%%%
%%%%%Ï%%%%%%%%%%%%%%%%%%%%%%%%%%%%%%%%%%%%%%%%%%%%%%%%%%%%%%%%%

%In the  a first 
%generation of unenriched low-mass stars forms along with the massive stars in a proportion expected
%from a normal IMF, 
%%(Salpeter + lognormal), 
%it has to be ejected almost completely,
%without loss of the enriched low mass stars. The latter implies mass segregation,
%the preferential formation of the enriched stars near the massive stars, 
%generally expected in self-enrichment scenarios, 
%and the expulsion of a dominant gaseous component
%on the crossing timescale 
%($\tau_\mathrm{c} = 0.2\,\mathrm{Myr}\,r_\mathrm{h,3}^{3/2} M_\mathrm{tot,6}^{-1/2}$,
%$r_\mathrm{h,3}$: half-mass radius in units of 3~pc, $M_\mathrm{tot,6}$: 
%total cluster mass in units of $10^6$~M$_\odot$),
%which would change the gravitational potential enough to lose the less bound 
%first generation low-mass stars. 
%
\subsection{How may gas disappear from a star cluster?}
In general, the remaining gas from which the stars in a cluster formed may
be cleared in three ways: First, star formation can proceed and use up the gas.
Second, a steady, moderately mass loaded wind \rvtwo{\citep[e.g.,][]{Palea13,Calurea15}} may 
remove gas over timescales
long compared to the crossing time, which is probably what happens in 
galactic winds \citep[e.g., ][]{SH09,vGlea13}. 
In this case, the change in gravitational potential is comparatively slow, such that 
the effect on the stellar kinematics is \rv{moderate, and the fraction of stars lost is small
\citep{BK07}. This would not be of interest in the present context.} 
Finally,
superbubble formation is a third possibility \citep[Paper~I in the following]{Krausea12a}.
Here, the hot gas is inside a thin shell, such that the majority of the gas in the thin
shell may be efficiently accelerated.
This is the most plausible setup for gas expulsion on the crossing timescale,
and thus has the potential for strong effects on the stellar kinematics.

The gas expulsion paradigm involves ejection of a gas mass comparable to the total mass 
on the crossing timescale ($\tau_\mathrm{c} = 0.2\,\mathrm{Myr}\,r_\mathrm{h,3}^{3/2} M_\mathrm{tot,6}^{-1/2}$,
$r_\mathrm{h,3}$: half-mass radius in units of 3~pc, $M_\mathrm{tot,6}$: 
total cluster mass in units of $10^6$~M$_\odot$). 
The binding energy due to the self-gravity of the gas is 
$10^{52} \,\mathrm{erg}\,(1-\epsilon_\mathrm{SF})M_\mathrm{tot,6}^{2} r_\mathrm{h,3}^{-1}$, where we have used the formula for a Plummer star cluster model \citep{BKP08}.
\rv{ $\epsilon_\mathrm{SF}$ is the star formation efficiency. Here, we mean a local
star formation efficiency, namely the initial stellar mass of a cluster divided 
by the total baryonic mass (stars and gas) within the space occupied by 
the forming stellar population of a given star cluster. }
The binding energy thus approaches the energy that supernovae can produce
within a crossing time \citep[compare][]{Decrea10}. 
In order to \rv{still expel the gas in these circumstances,} 
a very efficient gas expulsion mechanism 
is required.  A superbubble is therefore a promising candidate 
to fulfil the constraints. 
%Here, we study the evolution of a superbubble
%in a Plummer sphere for stars and gas.

\subsection{Energy production}
Stellar winds, ionising radiation and radiation pressure 
from massive stars appear very early 
in the formation of a star cluster \citep[compare, e.g.,][]{Dalea15b}.
After a few Myr, stellar explosions comprise the main energy source, and after a given star
has exploded, it may release energy by accretion, either from a companion
star or, for favourable conditions, from the intracluster medium.
\rvtwo{Technically, we derive the energy production rate by integrating the stellar
sources over the IMF, similar to, e.g., \citet{ShulSak95}. 
In a solar-metallicity star cluster, 
winds and supernovae combine to maintain a similar energy input rate over tens
of Myr \citep{Vossea09}.}

We consider the following cases for the energy production:

\begin{enumerate}

\item Stellar winds with the same energy injection rates as in \citetalias{Krausea12a},
but scaled for metallicity $Z$ as $Z^{0.7}$ \citep{MM12}, according to 
the metallicity of each cluster. We show in appendix~\ref{ap:rp_vs_thermal}
that the gas pressure due to thermalised stellar winds is much larger
than the radiation pressure and that radiation pressure alone may hardly ever
lead to gas expulsion in massive star clusters. 
Therefore, we neglect the radiation pressure. Effects of ionisation are also
negligible in the present context, because the escape speed in massive star clusters
(compare below) is much larger than the typical sound speed in photo-ionised 
gas \citep[compare, e.g.,][]{Dalea14a}.

\item Supernovae explosions of the massive stars at the end of their evolution. 
\rv{We assume that all massive stars explode as SNe, although this is subject to caution,
because stars more massive than $\approx 25$ \msun~might in fact silently turn into black holes.}
For the standard case, we assume an energy of $10^{51}$~erg per explosion.
Such explosions may, however, be enhanced by rapid spin-down, if a fast rotating
dark remnant was produced \citep[\rv{"hypernova", }e.g.,][]{Langer12}. The energy that may be released in this way
is limited by the extractable energy of black holes of mass $M_\mathrm{DM}$, 
\citep{Christo70,Hawking72,Penrose72,BZ77}, 
$E_\mathrm{x}=0.29 M_\mathrm{DM}c^2=5 \times 10^{54} \,\mathrm{erg} \,M_\mathrm{DM}/(10M_\odot)$.
Neutron stars will in general have lower rotational energies than this. Constraints from the spin-down
power of observed young pulsars, $\gtrsim 10^{38}$~erg~s$^{-1}$ for $\gtrsim10^3$~years
\citep{KH15}, provide a lower limit to the potential total energy release 
of $3 x 10^{48} \,\mathrm{erg}$. Observational constraints for rotating black hole-related
processes may be taken from measurements of type~Ic supernovae associated with GRBs
\citep[e.g.,][]{Mazea14}, which are around $10^{52}$~erg. \citet{Heesea15} find 
$\gtrsim10^{52}$~erg in a superbubble around the $>23M_\odot$ black hole IC10-X1,
plausibly related to a hypernova at its formation. 
\rv{If the strong explosions would be connected to \rvtwo{a} particular range of stellar
masses, they might occur in a small time interval. In the context of
gas expulsion, only the energy produced in a crossing time matters.
Hence, it would be possible that all the strong explosions in a given cluster
add up coherently to enable the gas expulsion process.}
We consider \rv{hypernovae with, individually, 10 and 100 times the standard
explosion energy of $10^{51}$~erg.} For a given simulation, all explosions have the same energy.
%\item Winds plus ionising radiation: The ionising luminosity is assumed to
%be trapped by the dense gas inside the star cluster. The power is added to the 
%stellar winds (always solar metallicity). We follow \citet{Marksea12} in using the fit of \citet{BKP08} to the
%radiation hydrodynamics simulations of \citet{FHY03,FHY06}:
%\mbox{log$_{10}	(E_* /\mathrm{erg\,	Myr}^{-1})= 50.0 + 1.72\sbr{\mathrm{log}_{10} (m /\mathrm{M}_\odot)- 1.55}$}, where $E_*$ denotes the energy output of a single star
%and $m$ is the mass of the star.
%\item Coherent dark remnant accretion as in \citetalias{Krausea12a}, i.e. after 
%all massive stars have reached the end of their nuclear burning and the dark remnants
%have been formed, accretion on to all the black holes
%(formed by all stars with initial mass $>25 M_\odot$), and optionally also the neutron stars,
%sets on coherently at the Eddington rate.
\end{enumerate}

\subsection{Implementation}\label{s:m:imp}

We evolve the dynamical equations for the expansion of
a spherically symmetric thin shell under conservation of energy and momentum,
\eql{eq:tse}{ \frac{\partial}{\partial t} ({\cal M} v) = p A - {\cal M}
  g\, ,}
where, ${\cal M}= 4\pi \int_0^r \rho_\mathrm{g}(r^\prime) r^{\prime\,2} \,\mathrm{d}r^\prime$ 
is the mass in the shell, with the gas density $\rho_\mathrm{g}$.
\rv{Shell radius and velocity are respectively denoted by} $r$ and $v$. 
\rv{The shell's surface area is
given by} $A=4\pi r^2$, and $g$ \rv{is the} gravitational acceleration.
The bubble pressure is $p=(\gamma-1) (\eta E(t)-{\cal
  M}v^2/2)/V$, with the bubble volume $V=4\pi r^3/3$, the energy
injection law $E(t)$, an efficiency parameter $\eta$, and the ratio of
specific heats, $\gamma=5/3$. \rv{We assume $p$ to dominate
over the ambient pressure.}
In this approach, 
\rvtwo{following essentially \citet{BBT95},}
the gravity of
the stars as well as the self-gravity of the gas are taken into account.
Detailed 3D hydrodynamics simulations, which take into account heating
and cooling processes in the gas and model instabilities in detail
yield a time-averaged radiative dissipation in supershells of about 
90 per cent of the injected energy \citep{Krausea13a,KD14}.
We therefore conservatively assumed that 80 per cent of the injected energy 
is radiated and 20 per cent (our parameter $\eta$) is used to move the gas.
We \rv{first} assumed a star formation efficiency $\epsilon_\mathrm{SF}$ of 30 per cent for 
some standard runs, because this is a reference value in gas expulsion studies
\citep[e.g.,][]{Decrea10},
but var\rv{ied} this parameter frequently as indicated in the individual investigations.
We use the \rvtwo{IMF} from \citet{Kroupea13}.
The initial condition is assumed to be a Plummer sphere for stars and gas.

Because gravity is strongest at the half-mass radius
the general form of the solution is a comparatively slow push across the half-mass
radius, and then an acceleration thanks to the decreasing gravitational force. 
\rv{Unless the shell's deceleration exceeds gravity}, it is prone to the 
Rayleigh-Taylor instability. We \rv{calculate} analytically when modes comparable
to the shell size have had enough time to grow.
\rv{The Rayleigh-Taylor length scale $\lambda$ is given by 
$\lambda =(a-g)\tau^2$, where $a$ is the shell's acceleration, and $\tau$ the time interval for which the instability 
criterion, $a-g>0$, was fulfilled \citepalias[compare][]{Krausea12a}.
\rvtwo{Note that a shell is always unstable if $a>0$ (shell acceleration), because
gravity is attractive ($g<0$). It is also unstable if the shell decelerates, but the 
magnitude of gravity exceeds the deceleration ($|g|>|a|$). Stability requires deceleration and 
small gravity ($|g|<|a|$).}

When $\lambda$ exceeds the shell radius, 
}
the flow is expected
to change character from a superbubble into a convective flow.
\rvtwo{This is a reasonable assumption:
The timescale $\tau$ defined above is very similar to the growth time
for a mixing layer \citep{PoPey10}, and also simply the time a piece of shell
material would need to reach the centre, if gravity dominates. 
Simulations show directly that smaller scale modes, for example ones comparable
to the shell thickness, lead to minor diffusion of shell material into the bubble interior
and bulk acceleration of the possibly fragmented shell
\citep[e.g.,][]{Krause2005b,Krausea13a}, whereas modes comparable to the bubble size
lead to large-scale overturn \citep[e.g.,][]{JM96,HK13}.}

Any remaining mass loss is then considerably reduced.
We refer the interested reader to \citetalias{Krausea12a} for more details 
about the numerics and basic behaviour of the solution.
%%%%%%%%%%%%%%%%%%%%%%%%%%%%%%%%%%%%%%%%%%%%%%%%%%%%%%%%%%%%%%
% cluster 23
%%%%%%%%%%%%%%%%%%%%%%%%%%%%%%%%%%%%%%%%%%%%%%%%%%%%%%%%%%%%%%
\begin{figure}
  \centering
  \includegraphics[width=0.47\textwidth]{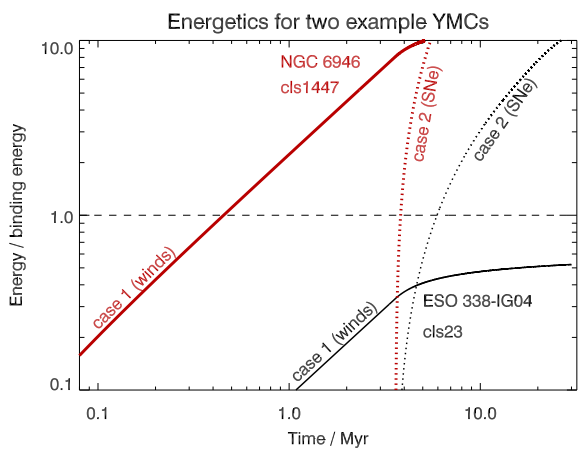}
  \caption{Produced energy in units of the binding energy 
        for a model with $\epsilon_\mathrm{sf}=0.3$ for \rv{two example
	clusters, ESO~338-IG04 cluster~23 (thinner black lines) and 
	NGC~6946 cluster~1447 (thicker red lines),} for different assumptions
        about the energy production: \rv{case 1 assumes stellar winds (solid lines); 
	case 2 is for standard supernovae (dotted lines). The dashed} 
	black horizontal line indicates the binding
	energy. 
       }
   \label{f:cl23e}%
\end{figure}
%%%%%%%%%%%%%%%%%%%%%%%%%%%%%%%%%%%%%%%%%%%%%%%%%%%%%%%%%%%%%%

\section{Results}\label{s:results}
%%%%%%%%%%%%%%%%%%%%%%%%%%%%%%%%%%%%%%%%%%%%%%%%%%%%%%%%%%%%%%
%%%%%%%%%%%%%%%%%%%%%%%%%%%%%%%%%%%%%%%%%%%%%%%%%%%%%%%%%%%%%%
\subsection{Young massive clusters}\label{s:YMCs}
%%%%%%%%%%%%%%%%%%%%%%%%%%%%%%%%%%%%%%%%%%%%%%%%%%%%%%%%%%%%%%
%%%%%%%%%%%%%%%%%%%%%%%%%%%%%%%%%%%%%%%%%%%%%%%%%%%%%%%%%%%%%%

We first present results for the sample of YMCs recently compiled by \citet{Bastea14b}
for the purpose of demonstrating the absence of gas in massive star clusters of ages 
less then or just about 10~Myr. For the calculations, we adopt the masses and metallicities
given in \citet{Bastea14b} (Table~\ref{t:YMCs}). 
We calculated the half-mass radius from the half-light
radii given in \citet{Bastea14b} by multiplying 1.7 which would be correct 
if the clusters were all Plummer spheres.

%%%%%%%%%%%%%%%%%%%%%%%%%%%%%%%%%%%%%%%%%%%%%%%%%%%%%%%%%%%%%%
\rv{\subsubsection{Cluster 1447 in the galaxy NGC~6946: \newline gas expulsion for standard conditions}\label{s:cl1447}}
%%%%%%%%%%%%%%%%%%%%%%%%%%%%%%%%%%%%%%%%%%%%%%%%%%%%%%%%%%%%%%

\rv{As an example for a star cluster where gas expulsion would be feasible,
we discuss the case of cluster~1447 in the galaxy NGC~6946. 
It has a stellar mass of $8\times 10^5 M_\odot$ and a comparatively 
large half-mass radius of 17~pc. It is gas-free at an age of 12~Myr.
The energy tracks are shown in Fig.~\ref{f:cl23e} (red lines). Stellar winds and supernovae
are able individually to provide an energy comparable to the binding energy within less 
than a Myr.
}

\begin{figure*}
  \centering
  \includegraphics[width=0.49\textwidth]{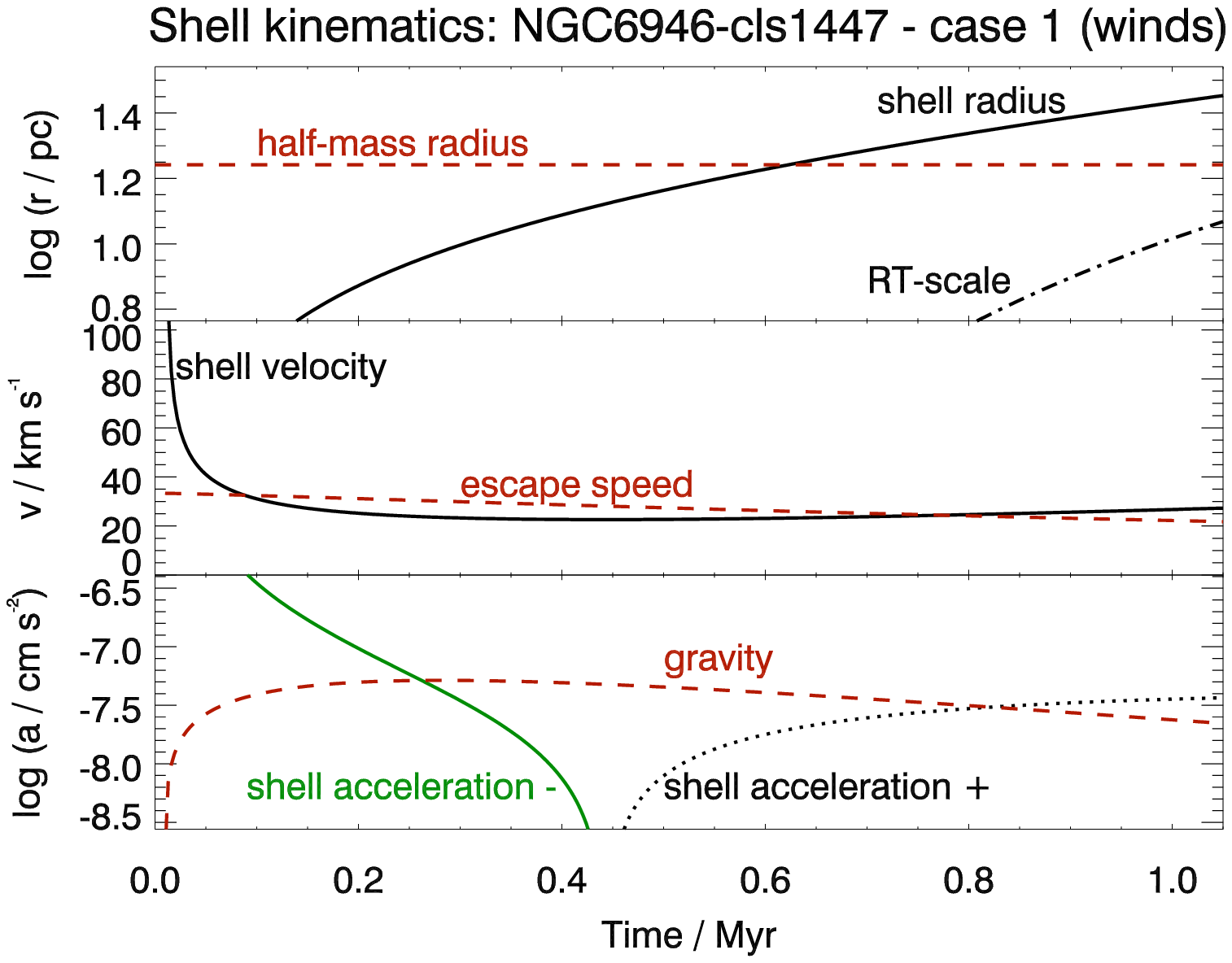}
  \includegraphics[width=0.49\textwidth]{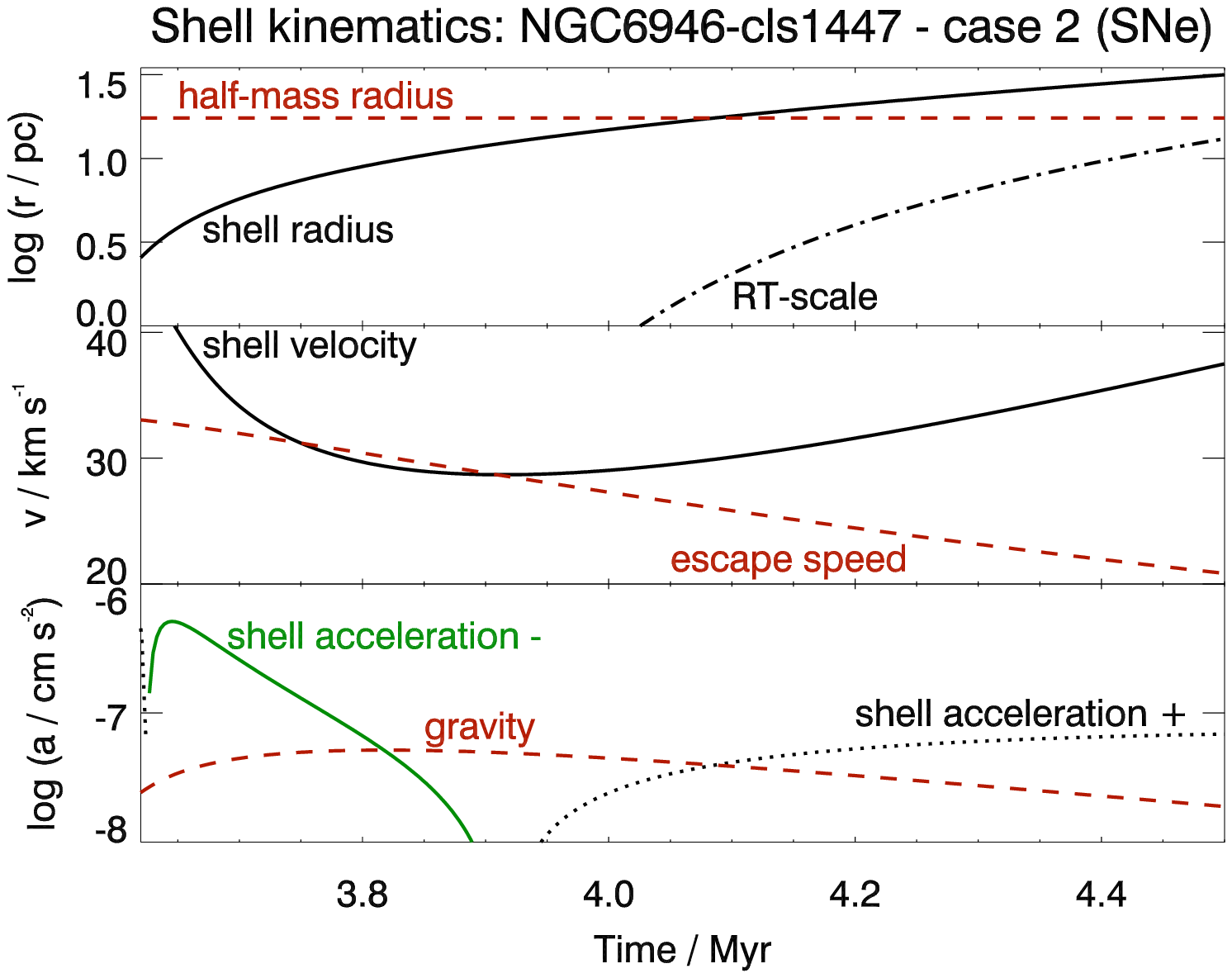}
  \caption{\rv{Supershell kinematics
        for cluster~1447 in the galaxy NGC~6946. 
	Left: case~1, stellar winds. Right: case~2, supernovae.
	The star formation efficiency is assumed to be 30~per cent in both cases.
        The timescale for the global evolution of the GC is chosen
        at the birth of a coeval first generation of stars. 
        Within each
        kinematics plot, the upper diagram shows the  bubble radius
        (solid black line) and the Rayleigh-Taylor scale (dash-dotted black line),
        with the red dashed line indicating the half-mass radius. The
        middle diagram displays the shell velocity (solid line) and
        the escape velocity at the current bubble radius (red dashed
        line). The acceleration is shown in the bottom diagram
	(positive: dotted black line, negative:
        solid green line), with the
        gravitational acceleration at the current radius shown as a red
        dashed line.}
	\rv{Gas expulsion succeeds in both cases. Details in Sect.~\ref{s:cl1447}}
	}
   \label{f:cl1447}%
\end{figure*}

\rv{We show the shell kinematics plot in Fig.~\ref{f:cl1447}. For both cases
(winds left, supernovae right), the Rayleigh-Taylor scale stays comfortably
below the shell radius at all times of interest (top diagrams). This means that only small
scale Rayleigh-Taylor modes disturb the shell, but overall the shell is preserved.
Only for short time intervals around the peak of the gravitational pull 
(bottom diagrams), the shell velocity
drops below the local escape speed (middle diagrams). But since gravity and shell
acceleration are too small to make the shell entirely unstable, the shell may
reaccelerate as a whole, beyond escape speed. After crossing the half-mass radius,
the gravitational pull declines, and the shell consequently accelerates (bottom diagrams).
The shell would then continue to accelerate indefinitely, because of the strong
decline of the density in the Plummer potential \citepalias[compare][]{Krausea12a}.
In reality it would soon be slowed down again by interaction with the ambient medium.}

\rv{Therefore in that peculiar case, our model with standard assumptions for the SFE and the sources of energy predicts a successful gas expulsion, consistent with observations.\\}

%%%%%%%%%%%%%%%%%%%%%%%%%%%%%%%%%%%%%%%%%%%%%%%%%%%%%%%%%%%%%%
\subsubsection{Cluster 23 in the galaxy ESO 338-IG04\rv{: \newline no gas expulsion for standard conditions}}\label{s:cl23}
%%%%%%%%%%%%%%%%%%%%%%%%%%%%%%%%%%%%%%%%%%%%%%%%%%%%%%%%%%%%%%

\rv{C}luster~23 in \object{ESO 338-IG04} has a mass of about 
$5\times 10^6 M_\odot$ and a half-mass radius 
of 9~pc. We \rv{again} assumed a star formation
efficiency of $\epsilon_\mathrm{SF}=0.3$.

\rv{T}he evolution of the energy production in this cluster \rv{is also shown in} 
Fig.\ref{f:cl23e} \rv{(black lines)}.
Stellar winds alone do not reach the binding energy for this cluster.
For \rv{supernovae},  the binding energy is reached only after about
1~Myr \rv{of supernova activity}. 

\begin{figure*}
  \centering
  \includegraphics[width=0.49\textwidth]{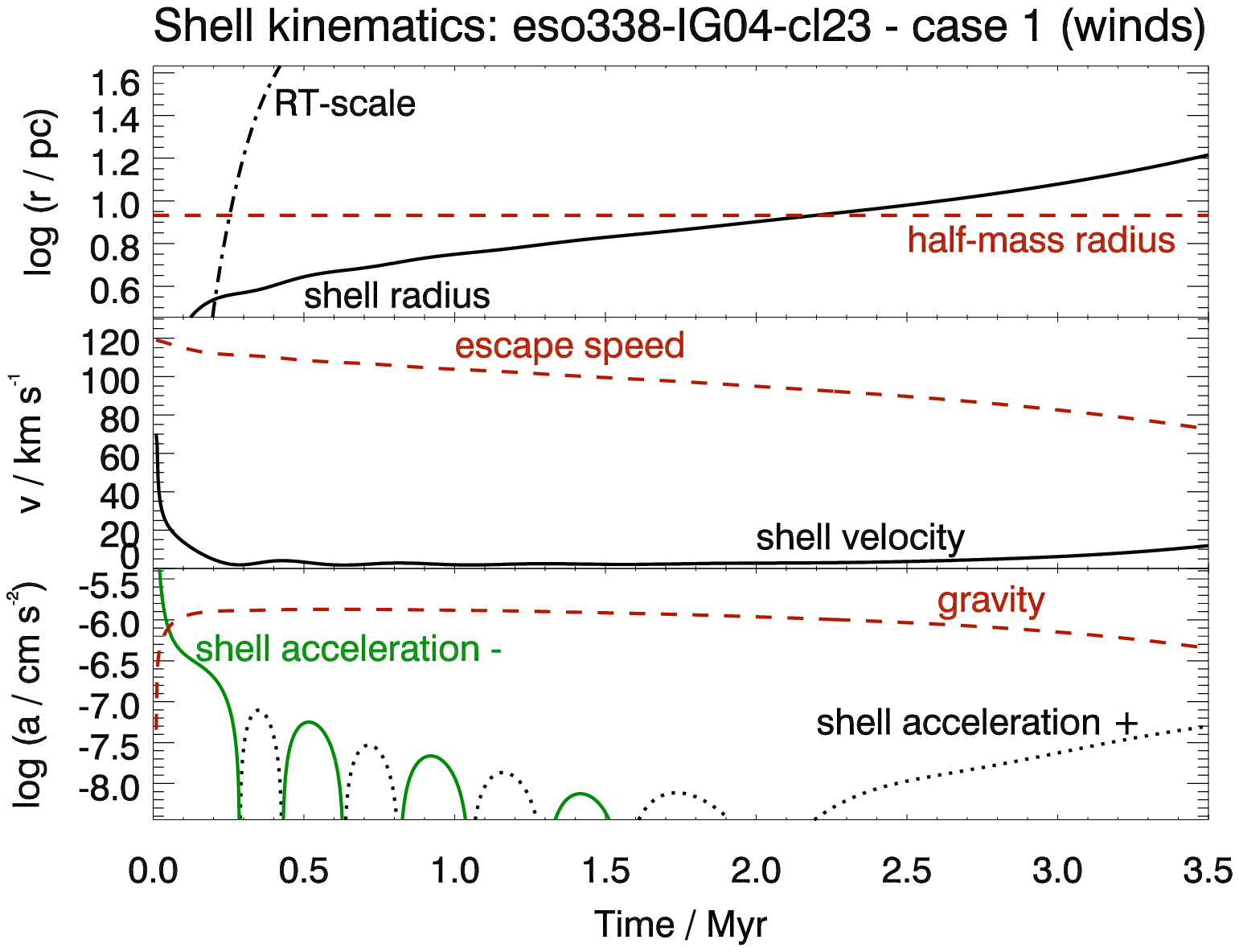}
  \includegraphics[width=0.49\textwidth]{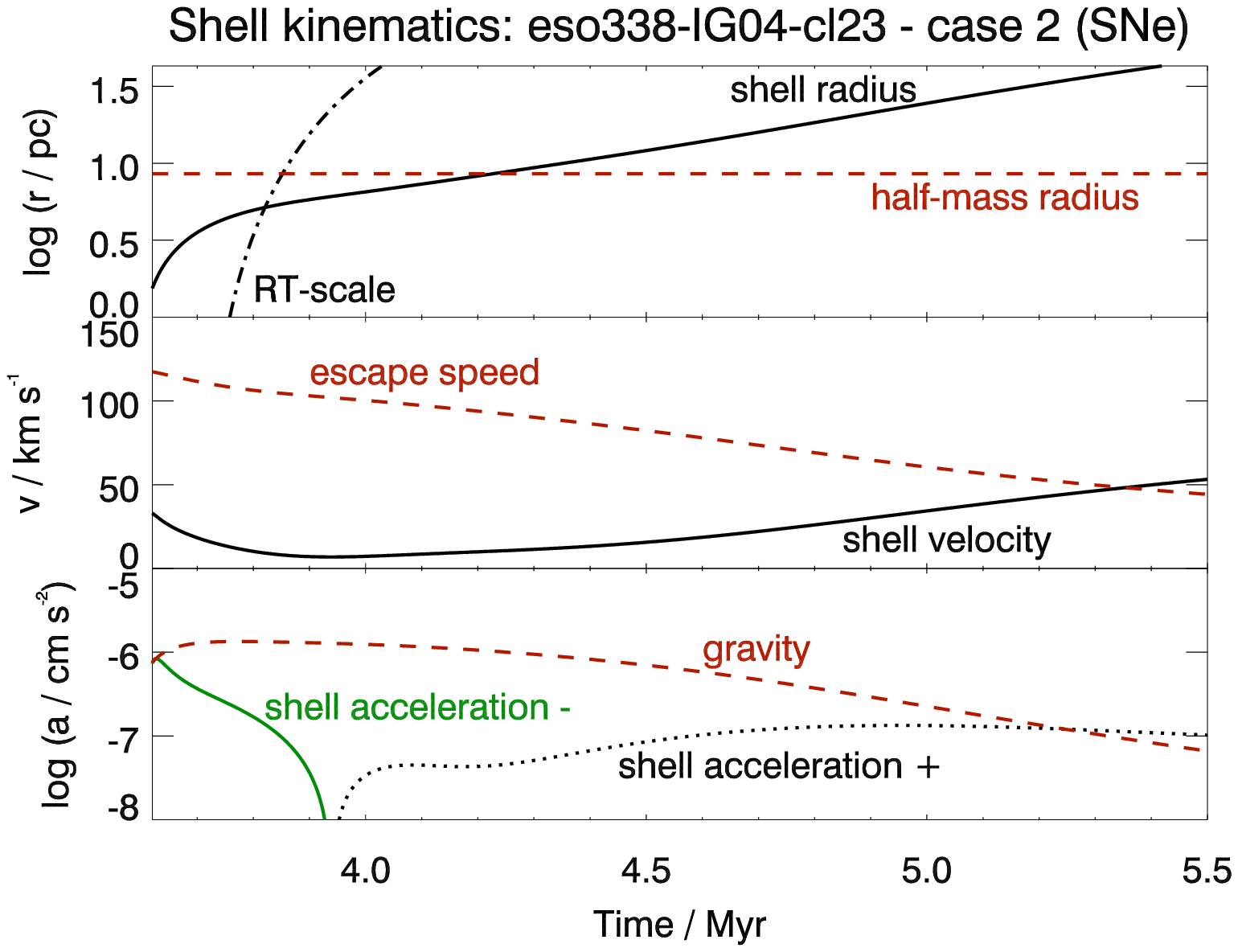}
  \caption{\rv{Same as Fig.~\ref{f:cl1447}, but for cluster~23 in the galaxy 
	\object{ESO338-IG04}. Gas expulsion does not succeed here. Details in 
	Sect.~\ref{s:cl23} }}
   \label{f:cl23}%
\end{figure*}

The energetics are reflected in the shell kinematics plots shown in Fig.\ref{f:cl23}. 
As expected, stellar winds (\rv{left} plot)
alone cannot expel the gas in this YMC for the adopted star formation efficiency\rv{.
This is also evident from the fact that the shell velocity is unable to reach escape speed
for the entire wind phase, up to about 3.5~Myr (middle diagram).
The shell is pushed so gently across the half-mass radius that the Rayleigh-Taylor
scale quickly exceeds the shell radius (top diagram). The shell is then destroyed
by the instability, and most of the gas \rv{is expected to fall} back (fountain flow). This is similar to
the interstellar medium in disc galaxies \citep[compare, e.g., ][]{vGlea13}.
The supershells studied here have a quite standard early deceleration phase (green line, bottom diagram). The deceleration, however, very quickly drops below gravity
(red dashed line, bottom diagram). Consequently, it cannot stabilise the shell.}

Supernovae \rv{(Fig.~\ref{f:cl23}, right)} could accelerate the supershell to escape speed 
within about 2~Myr 
(5.\rv{4}~Myr after the assumed coeval starburst
\rv{middle diagram).} However, this solution is \rv{also}
Rayleigh-Taylor unstable
because \rv{the shell \rvtwo{de}celeration still drops too quickly below the gravitational
acceleration. Gravity then boosts the Rayleigh-Taylor scale. It exceeds the shell radius 
from about 0.2 Myr after onset of the supernovae. Again, the shell is destroyed.
Similar to the wind phase, the intracluster medium will then be dominated by blowout of
individual bubbles \citep{Tenea15}.}

Therefore, if the star formation efficiency had been indeed 30 per cent, our calculations
would predict that the unused gas could not have been expelled within the first few Myr
by stellar winds, or standard supernovae. The ISM in this star cluster should then
be convective, similar to the ISM of the Milky Way and other star forming galaxies, where
the stellar (and other) energy sources lead to turbulence with typical velocities of the
order of 10~\kms. Because this is small compared to the escape velocity, the gas 
\rvtwo{would} remain bound. However, the cluster is \rvtwo{observed to be} gas-free at an
estimated age between 4~and 10~Myr.
%%%%%%%%%%%%%%%%%%%%%%%%%%%%%%%%%%%%%%%%%%%%%%%%%%%%%%%%%%%%%%
% cluster 1447
%%%%%%%%%%%%%%%%%%%%%%%%%%%%%%%%%%%%%%%%%%%%%%%%%%%%%%%%%%%%%%
%%%%%%%%%%%%%%%%%%%%%%%%%%%%%%%%%%%%%%%%%%%%%%%%%%%%%%%%%%%%%%
%%%%%%%%%%%%%%%%%%%%%%%%%%%%%%%%%%%%%%%%%%%%%%%%%%%%%%%%%%%%%%
% cluster 23
%%%%%%%%%%%%%%%%%%%%%%%%%%%%%%%%%%%%%%%%%%%%%%%%%%%%%%%%%%%%%%
%%%%%%%%%%%%%%%%%%%%%%%%%%%%%%%%%%%%%%%%%%%%%%%%%%%%%%%%%%%%%%
%%%%%%%%%%%%%%%%%%%%%%%%%%%%%%%%%%%%%%%%%%%%%%%%%%%%%%%%%%%%%%
% Table: SFE / the number is the last SFE, where gas removal did not work
%%%%%%%%%%%%%%%%%%%%%%%%%%%%%%%%%%%%%%%%%%%%%%%%%%%%%%%%%%%%%%
\begin{table*}
\caption{\rv{Observational data and results of gas expulsion simulations}
 for the YMC sample of  \citet{Bastea14b}.}
\label{t:YMCs}      % is used to refer this table in the text
\centering                          % used for centering table
\begin{tabular}{llccccccccc}        % centered columns (4 columns)
\hline\hline                 % inserts double horizontal lines
Galaxy & Cluster & Age\tablefootmark{a} & $M_*$\tablefootmark{b}
	& $r_\mathrm{h}\tablefootmark{c}$ & $C_5$\tablefootmark{d} & $Z$\tablefootmark{e} 
	& Ex/ & Ex/  
	& $\epsilon_\mathrm{SF,W,c}$\tablefootmark{g} & $t_\mathrm{ex,w}$\tablefootmark{h}\\
 	& & (Myr) & $(10^5\msm)$ & (pc) &  & ($Z_\odot$) & W\tablefootmark{f} & SN\tablefootmark{f} 
	&(\%)& (Myr) \\
\hline
\object{NGC 6946}      & 1447                & $12\pm2.5$   & 8   &  17.4 & 0.46 & 0.5 & 
	Y  & Y  & 20 & 0.63\\
\object{NGC 1569}      & A                      & $6\pm1$        & 7.6 & 1.5  & 5.1 & 0.4 & 
	N & N  &  80 & 0.12 \\
                                   & B                      & $15\pm5$       &14   & 2.4 &  5.9 & 0.4 &  
	N & N  & 80 & 0.13 \\
\object{NGC 1705}      & 1                      & $12.5\pm2.5$ &11  & 1.5  &  7.3 & 0.33 & 
	N & N  &  80 & 0.12 \\
\object{NGC 1140}      & 1                      & $5\pm 1$        &11  & 14   &  0.79 & 0.5 &  
	Y & Y  & 30 & 0.48\\
\object{The Antennae} & T352/W38220 & $4\pm2$         & 9.2 & 4.1 & 2.2 & 1 & 
	N & N & 40 & 0.20 \\ 
                                   & Knot S              &  $5\pm1$        &16   & 14  & 1.1 & 1 & 
	N & Y  & 40 & 0.52 \\
\object{ESO 338-IG04}& Cluster 23        &  $6^{+4}_{-2}$   & 50  & 8.9 & 5.6 & 0.2 & 
	N & N  & 80 & 0.22 \\
\hline                                   %inserts single line
\end{tabular}
\tablefoot{ 
\tablefoottext{a}{Age in Myr.}
\tablefoottext{b}{Fiducial stellar mass, adopting the photometric mass estimate from \citet{Bastea14b}.}
\tablefoottext{c}{$r_\mathrm{h}$: half-mass radius, calculated from the half-light radius assuming a Plummer sphere.}
\tablefoottext{d}{Compactness index $C_5 = (M_*/10^5\msm) (r_\mathrm{h}/\mathrm{pc})^{-1}$}
\tablefoottext{e}{Metallicity.}
\tablefoottext{f}{Success of gas expulsion assuming 30 per cent star formation efficiency
by, respectively, stellar winds (Ex/W) and
supernovae (Ex/SN). Y: yes, N: no.}
\tablefoottext{g}{Critical star formation efficiency for gas expulsion in the metallicity dependent massive-star wind scenario. With this efficiency, a supershell would just not yet escape. The simulation
at a star formation efficiency ten percentiles higher then features successful gas expulsion.} 
\tablefootmark{h}{Time since the coeval starburst at which an eventually escaping shell crosses
the half-mass radius, for the stellar wind case with $\epsilon_\mathrm{SF,W,c}$.}

}
\end{table*}
%%%%%%%%%%%%%%%%%%%%%%%%%%%%%%%%%%%%%%%%%%%%%%%%%%%%%%%%%%%%%%

%%%%%%%%%%%%%%%%%%%%%%%%%%%%%%%%%%%%%%%%%%%%%%%%%%%%%%%%%%%%%%
\subsubsection{Other YMCs}
%%%%%%%%%%%%%%%%%%%%%%%%%%%%%%%%%%%%%%%%%%%%%%%%%%%%%%%%%%%%%%
We have performed similar calculations 
%as for cluster~23 in ESO 338-IG04 
for all star clusters
of the sample of \citet{Bastea14b}. The results are summarised in Table~\ref{t:YMCs}.
For two of the objects, gas expulsion is predicted to be possible with a star formation 
efficiency of 30 per cent for \rv{both} energy injection cases studied 
(winds \rv{and} supernovae). For one cluster, stellar winds 
would have been insufficient, but supernovae would
succeed. For five objects, gas expulsion is neither possible by winds nor supernovae.

For each cluster, we performed simulations varying the star formation efficiency by adding 
a corresponding amount of initial gas to the observed stellar mass. 
From this we determined the critical star formation efficiency,
$\epsilon_\mathrm{SF,W,c}$,
at which the supershell would just not yet escape when driven by the stellar winds.
It is also given in Table~\ref{t:YMCs}. We find values as high as 80 per cent.

If the star formation efficiency would indeed rise until stellar winds are able to clear
the remaining gas, this would then happen on a timescale of a few $10^5$ yrs 
(Table~\ref{t:YMCs}), which would agree with the observational finding that these clusters
are gas-free after a few Myr.

%%%%%%%%%%%%%%%%%%%%%%%%%%%%%%%%%%%%%%%%%%%%%%%%%%%%%%%%%%%%%%
%%%%%%%%%%%%%%%%%%%%%%%%%%%%%%%%%%%%%%%%%%%%%%%%%%%%%%%%%%%%%%
\subsection{Model grid for all cases of energy injection}\label{s:grid}
%%%%%%%%%%%%%%%%%%%%%%%%%%%%%%%%%%%%%%%%%%%%%%%%%%%%%%%%%%%%%%
%%%%%%%%%%%%%%%%%%%%%%%%%%%%%%%%%%%%%%%%%%%%%%%%%%%%%%%%%%%%%%

We run a model grid, for stellar masses of $10^5$, $10^6$, and $10^7$~\msun,
and half-mass radii of 1, 3, and 10~pc, varying the star formation efficiency
in steps of $\Delta\epsilon_\mathrm{SF}=0.1$ for the following energy 
production models: stellar winds at low metallicity (\feh{-1.5}), stellar winds
at solar metallicity, normal supernovae, hypernovae at $10^{52}$~erg each, \rv{and} 
hypernovae at $10^{53}$~erg each. For each scenario,
we determined the critical star formation efficiency for gas expulsion, and,
where the grid contained cases of successful gas expulsion, the time for the
supershell to reach the half-mass radius. 
We define the critical star formation efficiency for gas expulsion
$\epsilon_\mathrm{SF,crit}$ as the one where gas expulsion does just not yet
succeed. The model with the next higher star formation efficiency (by 0.1)
would then
feature successful gas expulsion.
The results are shown in Table~\ref{t:sfe-w-lz}.

The critical star formation efficiency for this model grid assumes the full range
of possible values. It generally increases for increasing mass and decreasing
radius. For the low metallicity winds, even the $10^5$~\msun~clusters have
critical star formation efficiencies above 30 to 80 per cent. For the $10^7$~\msun
clusters, it approaches 100 per cent. At solar metallicity, 
$\epsilon_\mathrm{SF,crit}$ drops to zero
for the lightest and most extended cluster, whereas the heavy and concentrated
ones still have values of 90 per cent. The table for normal supernovae
looks remarkably similar to the one for solar metallicity winds.
Each factor of ten in explosion energy changes the values by a similar amount
than the change in metallicity in the wind models. 
Hypernovae with $10^{53}$~erg, each, reduce $\epsilon_\mathrm{SF,crit}$
to zero for all $10^5$~\msun clusters and to 80 per cent for the most compact
$10^7$~\msun~object.
%The "winds plus ionisation" results are generally somewhat below the ones
%from the solar metallicity winds models, as expected, since the former
%are essentially enhanced solar metallicity winds.

We calculated the time for the shell to reach the half-mass radius as a proxy 
for the gas expulsion timescale. For models that feature
successful gas expulsion, this timescale never exceeds the crossing time by 
more than a factor of 2.5 and is usually below the crossing time.
Using a much higher star formation efficiency than the critical one, where
possible, reduces this timescale further.

%%%%%%%%%%%%%%%%%%%%%%%%%%%%%%%%%%%%%%%%%%%%%%%%%%%%%%%%%%%%%%
% Table: SFE / the number is the last SFE, where gas removal did not work
%%%%%%%%%%%%%%%%%%%%%%%%%%%%%%%%%%%%%%%%%%%%%%%%%%%%%%%%%%%%%%
\begin{table}
\caption{Critical star formation efficiency $\epsilon_\mathrm{SF,crit}$ for successful gas removal in a certain scenario. Given is $\epsilon_\mathrm{SF}$ of the last model
that did not result in successful gas expulsion, where $\epsilon_\mathrm{SF}$
was varied in steps of 0.1, or 0 if all investigated models lead to gas expulsion.
If the number is less than 0.9, we give the gas removal timescale at a star formation efficiency higher by $\Delta\epsilon_\mathrm{SF,crit}=0.1$ in units of the crossing time in brackets.}             
\label{t:sfe-w-lz}      % is used to refer this table in the text
\centering                          % used for centering table
\begin{tabular}{c l l l}        % centered columns (4 columns)
\hline\hline                 % inserts double horizontal lines
log$(M_*/\msm)$\tablefootmark{a}& $r_\mathrm{h}=1$~pc\tablefootmark{b} & 
$r_\mathrm{h}=3$~pc\tablefootmark{b} & $r_\mathrm{h}=10$~pc\tablefootmark{b}\\
\hline
& & & \\[.1mm]
\multicolumn{4}{c}{--- stellar winds at low metallicity, $\sbr{\mathrm{Fe/H}}=-1.5$ ---}\\
& & & \\[.1mm]
5 & 0.8 (1.45) & 0.5 (0.73) & 0.3 (0.37)\\
6 & 0.9 & 0.8 (1.88) & 0.6 (0.63)\\
7 & 0.9 & 0.9 & 0.9 \\
& & & \\[.1mm]
\multicolumn{4}{c}{--- stellar winds at solar metallicity ---}\\
& & & \\[.1mm]
5 & 0.4 (1.90) & 0.3 (0.59) &  0 (0.63) \\
6 & 0.9 & 0.8 (0.84) & 0.4 (0.46) \\
7 & 0.9 & 0.9 &  0.8 (0.63) \\
& & & \\[.1mm]
\multicolumn{4}{c}{--- normal supernovae, $E_0 = 10^{51}$~erg ---}\\
& & & \\[.1mm]
5 & 0.5 (0.66) & 0.2 (0.55) & 0 (0.54) \\
6 & 0.9 & 0.7 (0.52) & 0.4 (0.37) \\
7 & 0.9 & 0.9 & 0.8 (0.42) \\
& & & \\[.1mm]
\multicolumn{4}{c}{--- hypernovae, $E_0 = 10^{52}$~erg ---}\\
& & & \\[.1mm]
5	& 0.2 (0.64)	& 0 (	0.72)       & 0 (0.25)\\
6	& 0.7 (0.56)	& 0.3 (0.61)	& 0.1 (0.43)	\\
7	& 0.9	        & 0.8 (0.52)	& 0.4 (0.55)	\\
& & & \\[.1mm]
\multicolumn{4}{c}{--- hypernovae, $E_0 = 10^{53}$~erg ---}\\
& & & \\[.1mm]
5	& 0 (0.80)	& 0 (0.31)	& 0 (0.11)	\\
6	& 0.3 (0.64)	& 0.1 (0.56)	& 0 (0.36)	\\
7	& 0.8 (0.67)	& 0.5 (0.55) & 0.1 (0.65)	\\
%
%& & & \\[.1mm]
%\multicolumn{4}{c}{--- winds $+$ ionisation ---}\\
%& & & \\[.1mm]
%%
%5 & 0.2 (2.52) & 0.1 (0.89) & 0 (0.49)\\
%6 & 0.9 & 0.4 (1.41) & 0.2 (0.61)\\
%7 & 0.9 & 0.8 (2.54) & 0.6 (0.82)\\
\hline                                   %inserts single line
\end{tabular}
\tablefoot{ 
\tablefoottext{a}{Base 10 logarithm of the initial stellar mass of the cluster in solar masses.}
\tablefoottext{b}{$r_\mathrm{h}$: half-mass radius}
}
\end{table}
%%%%%%%%%%%%%%%%%%%%%%%%%%%%%%%%%%%%%%%%%%%%%%%%%%%%%%%%%%%%%%

\begin{figure}
  \centering
  \includegraphics[width=0.49\textwidth]{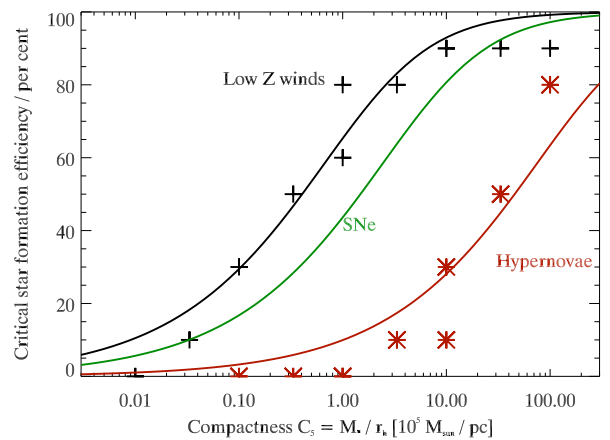}
  \caption{Critical star formation efficiency for gas expulsion $\epsilon_\mathrm{SF,crit}$
by low-metallicity stellar winds (black plus-signs) and $10^{53}$ erg hypernovae (red stars) 
	 versus compactness index $C_5$. The black and red solid lines
	are fits using equation~(\ref{eq:eps-scaling}). The fit to the models with normal supernovae is shown in green for comparison.}
   \label{f:comp-eps}%
\end{figure}

In a simple scaling argument, the binding energy of gas in a star cluster is 
proportional to 
$(1-\epsilon_{SF})M_\mathrm{tot}^2 / r_\mathrm{h}$ (compare Sect.~\ref{s:method}), where $M_\mathrm{tot}$ is the total initial mass. 
The feedback energy should scale with the stellar mass 
$M_* =\epsilon_\mathrm{SF} M_\mathrm{tot}$. Gas 
expulsion should happen when the supplied feedback energy reaches a certain threshold 
which should scale with the binding energy. This would imply
$(1-\epsilon_\mathrm{SF}) (M_*/\epsilon_\mathrm{SF})^2 / r_\mathrm{h} \propto M_*$, and hence: 
\eql{eq:eps-scaling}{
\epsilon_\mathrm{SF} = -\frac{aC_5}{2} + \sqrt{\frac{a^2 C_5^2}{4} + aC_5} \, ,
}
with the constant of proportionality $a$ and the 
compactness index $C_5 = (M_*/10^5\msm) (r_\mathrm{h}/\mathrm{pc})^{-1}$.
We show the critical star formation efficiency plotted against the compactness index
in Fig.~\ref{f:comp-eps}, together with fits of equation~(\ref{eq:eps-scaling}).
Clearly, equation~(\ref{eq:eps-scaling}) describes our models well.
We provide the values for the fit parameter $a$ for the respective cases in 
Table~\ref{t:a}.

%%%%%%%%%%%%%%%%%%%%%%%%%%%%%%%%%%%%%%%%%%%%%%%%%%%%%%%%%%%%%%
% Compactness versus critical star formation efficiency / Expulsion by winds
%%%%%%%%%%%%%%%%%%%%%%%%%%%%%%%%%%%%%%%%%%%%%%%%%%%%%%%%%%%%%%
%%%%%%%%%%%%%%%%%%%%%%%%%%%%%%%%%%%%%%%%%%%%%%%%%%%%%%%%%%%%%%

%%%%%%%%%%%%%%%%%%%%%%%%%%%%%%%%%%%%%%%%%%%%%%%%%%%%%%%%%%%%%%
% Table: Fit parameter a for eq (1)
%%%%%%%%%%%%%%%%%%%%%%%%%%%%%%%%%%%%%%%%%%%%%%%%%%%%%%%%%%%%%%
\begin{table}
\caption{Fit parameter $a$ of equation~(\ref{eq:eps-scaling}) for the different cases
for the energy production from Table~\ref{t:sfe-w-lz}.}             
\label{t:a}      % is used to refer this table in the text
\centering                          % used for centering table
\begin{tabular}{ll}        % centered columns (4 columns)
\hline\hline                 % inserts double horizontal lines
Case & ~~~~$a$\\
\hline
stellar winds at low metallicity, $\sbr{\mathrm{Fe/H}}=-1.5$ & 1.23 \\
stellar winds at solar metallicity & 0.368 \\
normal supernovae, $E_0 = 10^{51}$~erg & 0.337 \\
hypernovae, $E_0 = 10^{52}$~erg & 0.0554 \\
hypernovae, $E_0 = 10^{53}$~erg & 0.0110 \\
%winds $+$ ionisation & 0.109 \\
%
\hline                                   %inserts single line
\end{tabular}

\end{table}
\section{Discussion}\label{s:disc}
%%%%%%%%%%%%%%%%%%%%%%%%%%%%%%%%%%%%%%%%%%%%%%%%%%%%%%%%%%%%%%
%%%%%%%%%%%%%%%%%%%%%%%%%%%%%%%%%%%%%%%%%%%%%%%%%%%%%%%%%%%%%%

%%%%%%%%%%%%%%%%%%%%%%%%%%%%%%%%%%%%%%%%%%%%%%%%%%%%%%%%%%%%%%
% ESO338-IG04 Cl23 eps 90 / winds
%%%%%%%%%%%%%%%%%%%%%%%%%%%%%%%%%%%%%%%%%%%%%%%%%%%%%%%%%%%%%%
\begin{figure}
  \centering
  \includegraphics[width=0.50\textwidth]{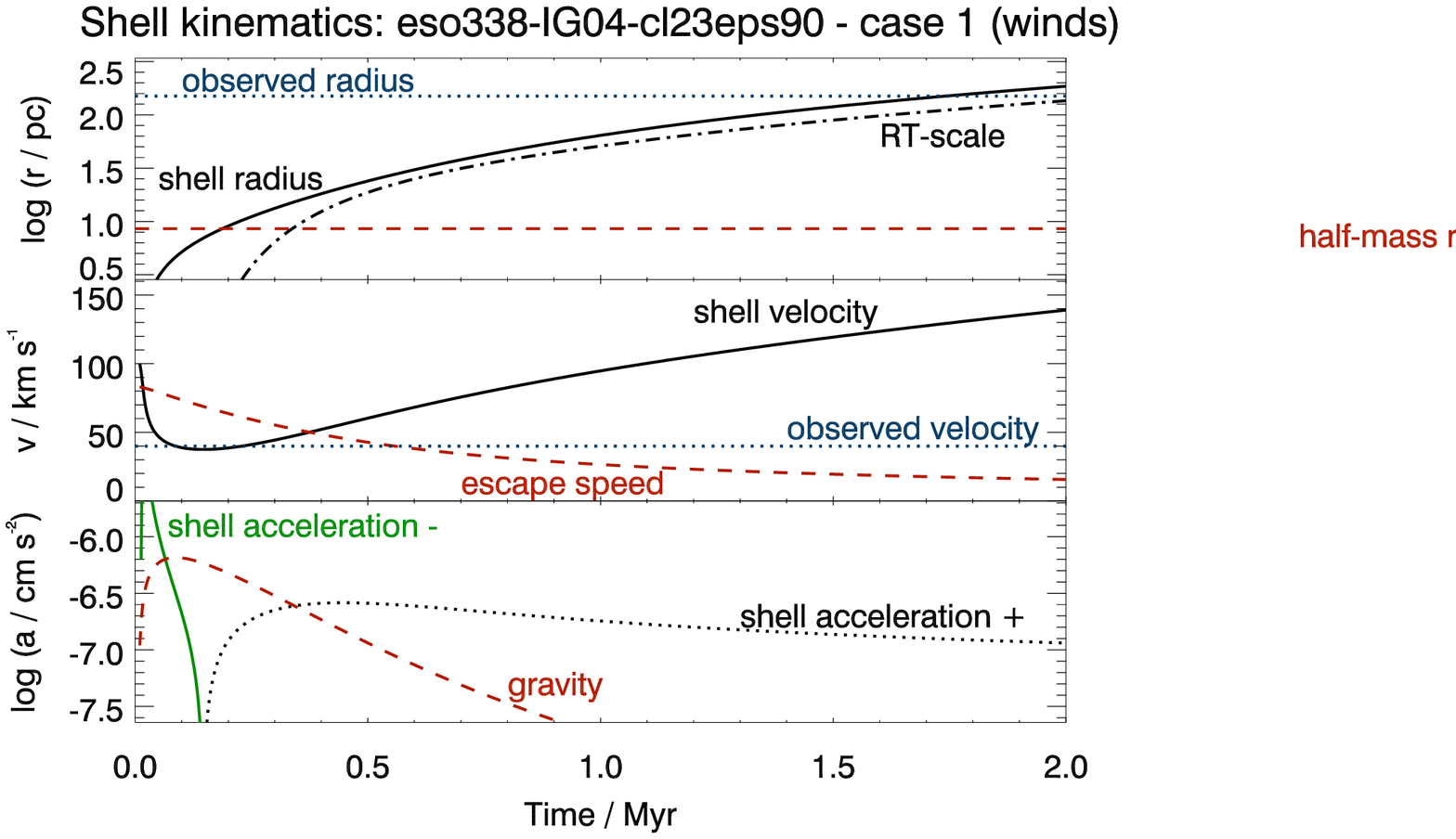}
  \caption{Shell kinematics for Cluster~23 in \object{ESO~338-IG04} for the 
	case where the energy is produced by stellar winds, and a star formation 
	efficiency of 90 per cent. The meaning of the 
	lines is the same as in Fig.~\ref{f:cl23}, except the blue dotted lines, 
	which give the observed radius (upper panel) and velocity (middle panel) 
	of the supershell around this object.
	The low observed velocity compared to our model would mean that most of the
	observed shell mass would be contributed by ISM around the cluster rather than 
	gas that was originally deep inside its potential well.}
   \label{f:cl23eps90-w}%
\end{figure}
%%%%%%%%%%%%%%%%%%%%%%%%%%%%%%%%%%%%%%%%%%%%%%%%%%%%%%%%%%%%%%
\subsection{Gas expulsion in young massive clusters} 
Five of the YMCs in the sample of \citet{Bastea14b}
could have expelled their residual gas neither by stellar winds nor by normal 
supernovae (and also not by radiation pressure, compare Sect.~\ref{s:method}),
if the star formation efficiency was 30 per cent, the preferred value for 
self-enrichment scenarios to explain the ratio of first to second generation stars.
Extrapolating our grid results
using equation (\ref{eq:eps-scaling}) with the parameter $a$ from Table~\ref{t:a},
we find that $10^{52}$~erg hypernovae would be needed to expel the gas 
at this star formation efficiency in one case (\object{T352/W38220}), and 
$10^{53}$~erg hypernovae would be required for the other four clusters.

The agent of the possible gas expulsion can
be constraint from the observationally inferred timescales:
Stellar explosions appear only after the most massive stars have completed
their main sequence and Wolf-Rayet phases, i.e. 3-4~Myr after formation.
The tightest constraints on timescales probably come from 
Cluster~23 in  \object{ESO~338-IG04} and Cluster~A in \object{NGC~1569}.
The former has an age of $6_{-2}^{+4}$~Myr. The surrounding superbubble
expands at $\approx40$~\kms and has a radius of $150\pm30$~pc
\citep{Bastea14b}. The inferred dynamical age of $3-4.5$~Myr implies
that any residual gas was expelled from the cluster $\approx0-5.5$~Myr 
after formation of the cluster.
Cluster~A in \object{NGC~1569} shares a superbubble with a somewhat older
neighbouring cluster, thus it could have lost its gas at any point during its lifetime
of $6\pm1$~Myr. For both objects, it would therefore still just be possible to lose the gas
at the beginning of the supernova phase, if the star formation efficiency was high, or if
the average energy per explosion was larger than $10^{51}$~erg~s$^{-1}$,
though stellar winds appear more likely from the timescale constraints.

One way to remove residual gas is indeed to assume a higher
star formation efficiency.
In Fig.~\ref{f:cl23eps90-w}, we show the shell kinematics for the successful
gas expulsion case by stellar winds, if the star formation efficiency was as high 
as 90 per cent. The shell would need 1.8~Myr to reach the observed radius, 
and the velocity would then be 130 ~\kms, about three times larger than observed.
However, the model assumes essentially zero density outside the cluster (Plummer 
model). The low observed velocity would then imply, unsurprisingly, that the shell
gas is mainly material that never fell deeply into the star cluster potential.
Thus, observations of the shell mass may only place lower limits on the 
star formation efficiency.

%Of course, late gas expulsion by coherent onset of dark remnant accretion,
%$\gtrsim 30$~Myr after star formation is excluded by the young age of these 
%clusters.

%%%%%%%%%%%%%%%%%%%%%%%%%%%%%%%%%%%%%%%%%%%%%%%
\subsection{Model grid: importance of compactness and star formation efficiency}
%%%%%%%%%%%%%%%%%%%%%%%%%%%%%%%%%%%%%%%%%%%%%%%
The model grids demonstrate that the compactness index, $C_5$, is the parameter
that governs gas expulsion. The numerical results, however, also show some real scatter, which is
most likely due to the non-linearity of the underlying physics. While the models shown in 
Fig.~\ref{f:comp-eps} generally follow equation~(\ref{eq:eps-scaling}) well, any two models with the 
same $C_5$ may still differ by 20 percentiles in critical star formation rate.

\rv{The critical star formation efficiency we derive could be interpreted
in the context of self-regulated star-cluster formation: 
once enough massive stars are formed,
the feedback energy would be sufficient to dispel the remaining gas and terminate
star formation. Such a process has been discussed by \citet{ZY07} for massive
star formation in general. Smooth particle hydrodynamics simulations of the 
formation of star clusters with feedback  find a more complex behaviour,
where gas clearance is sometimes dominated by accretion of the gas 
on to stars, but sometimes significant amounts of gas are driven away,
depending on the implementation of feedback \citep{Dale15}}.

\rv{Supernovae are probably not very important for gas expulsion at solar metallicities
or above: The injected power is comparable to that of the stellar winds.
The latter act, however, before the supernovae and are therefore more probable
to be the actual agents of gas removal, if it is removed.}

\rv{At low metallicity, supernovae could be important drivers of natal cloud destruction,
because, for a given local star formation efficiency, the supernovae inject more power.
 }

\rv{If star clusters hosting multiple hypernovae would exist, they would
of course be the dominant agents of gas removal. They would be able to 
cause gas expulsion (local star formation efficiency $< 50$ per cent)
up to a compactness index of $C_5 \approx 30$. This corresponds to, for example,
a $10^7 \msm$ star cluster with a half-mass radius of 3~pc.}

%%\rv{A}ssum\rv{ing,} for the sake of the argument\rv{,} that massive star clusters would 
%generally have an episode of gas expulsion, where the majority of stars were lost
%\rv{would} imply 
%star formation efficiencies \rv{around} 50 per cent \rv{\citep{KhalBaum15}.} 
%this would imply a narrow range of allowed compactness indices of about 1.5 dex (Fig.~\ref{f:comp-eps}).
%The actual $C_5$-values would be related to the feedback process actually realised in nature (also
%compare Sect.~\ref{s:dacc}). Measured $C_5$-values, however, spread over several orders of magnitude
%(Tables~\ref{t:YMCs} and~\ref{t:scdata} and Fig.\ref{f:c5-age}).

%\subsection{Gas expulsion in the progenitors of globular clusters}
%The original conditions before gas expulsion inferred by \citet{Marksea12} 
%for a sample of GCs imply very tightly bound star clusters. We show that 
%even with the strong winds with ionisation scenario, gas expulsion would not have succeeded
%in any one case once the Rayleigh-Taylor instability is taken into account. Even the
%extreme coherent dark remnant accretion scenario fails for many objects.
%The fact that gas expulsion seems impossible for some such objects is sufficient to demonstrate
%that this method of inferring the original star cluster parameters cannot be generally applicable.

%%%%%%%%%%%%%%%%%%%%%%%%%%%%%%%%%%%%%%%%%%%%%%%%%%%%%%%%%%%%%%
% Data for star clusters that have been searched for the anticorrelation
%%%%%%%%%%%%%%%%%%%%%%%%%%%%%%%%%%%%%%%%%%%%%%%%%%%%%%%%%%%%%%
\begin{table*}
\caption{Data for star clusters that have been searched for the Na-O anticorrelation \rv{or multiple
populations in the colour-magnitude diagram}}             
\label{t:scdata}      % is used to refer this table in the text
\centering                          % used for centering table
\begin{tabular}{l c c c c c c c c l}        % centered columns (4 columns)
\hline\hline                 % inserts double horizontal lines
\vspace*{.1mm}
Class & Object & Alt. name & 
$M_*$& $r_\mathrm{h}$\tablefootmark{a} & 
$C_5$\tablefootmark{b} & Age & $\sbr{\mathrm{Fe/H}}$ & MPs\tablefootmark{c} & Ref.\\[.2mm]
&&& $(10^5\msm)$ & (pc) &  &(Gyr)& & & \\
\hline
% all radii from Harris (half-mass). converted from arcmin to pc with distance from Harris
% from calc_carretta_rh.pro
GC & \object{NGC~~~104} & \object{47~Tuc}        & 6.46  & 7.1 & 0.92 & 12.8 & -0.76 & Y & 1, 2, 3, 4\\
GC & \object{NGC~~~288} & \object{Melotte~3}    & 0.46  & 9.8 & 0.05 & 12.2 & -1.32 & Y & 1, 2, 3, 4\\
GC & \object{NGC~~~362} & \object{Dunlop~62}   & 2.5   & 3.5 & 0.72 & 10 & -1.26 & Y & 2, 3, 4, 5\\
GC & \object{NGC~1261} & \object{Caldwell~87}   & 3.41   & 5.5 & 0.62 & 10.24 & -1.08 & Y & 2, 4, 6, 7\\
GC & \object{NGC~1851} & \object{Dunlop~508}   & 5.51   & 3.05 & 1.81 & 7.64 & -1.13 & Y & 2, 4, 6, 8\\
GC & \object{NGC~1904}   & M79     & 1.45 & 4.1  & 0.50 & 12.0 & -1.58 & Y & 1, 2, 9\\
GC & \object{NGC~2298}   & \object{Dunlop~578} & 0.85 & 3.1 & 0.16  & 12.4 & -1.98 & Y & 2, 4, 6, 8\\
GC & \object{NGC~2808}   & \object{Dunlop~265} & 12.3 & 3.8 & 3.24  & 11.2 & -1.18 & Y & 1, 2, 3, 4\\
GC & \object{NGC~3201}   & \object{Dunlop~445}           & 1.1   & 7.5 & 0.15  & 11.1 & -1.51 & Y & 1, 2, 4, 10\\
GC & \object{NGC~4590}   & M68     & 1.07 & 7.7 & 0.14  & 12.7 & -2.27 & Y & 1, 2, 3, 4\\
GC & \object{NGC~4833}   & \object{Dunlop~164}     & 4.10 & 7.9 & 0.52  & 12.5 & -1.71 & Y & 2, 4, 6, 7\\
GC & \object{NGC~5024}   & M53     & 8.26 & 11.6 & 0.71  & 12.7 & -1.86 & Y & 2, 4, 6, 7\\
GC & \object{NGC~5053}   &            & 1.25 & 22.5 & 0.056  & 12.3 & -1.98 & Y & 2, 4, 6, 7\\
GC & \object{NGC~5272}   & M3       & 4.68 & 11.7 & 0.40  & 11.4 & -1.34 & Y & 2, 3, 4, 7\\
GC & \object{NGC~5286}   &             & 7.13 & 4.2 & 1.69  & 12.5 & -1.41 & Y & 2, 3, 4, 7\\
GC & \object{NGC~5466}   &             & 1.79 & 18.2 & 0.098  & 13.6 & -2.20 & Y & 2, 4, 6, 7\\
GC & \object{NGC~5897}   &             & 2.11 & 12.7 & 0.17  & 12.3 & -1.73 & Y & 2, 4, 6, 7\\
GC & \object{NGC~5904}   & M5       & 3.89 & 6.6 & 0.59  & 11.5 & -1.33 & Y & 1, 2, 3, 4\\
GC & \object{NGC~5927}   &             & 3.38 & 4.2 & 0.81  & 12.7 & -0.64 & Y & 2, 4, 6, 7\\
GC & \object{NGC~5986}   &             & 5.99 & 5.04 & 1.19  & 12.2 & -1.35 & Y & 2, 4, 6, 7\\
GC & \object{NGC~6093}   & M80     & 5.02 & 3.02 & 1.66  & 12.5 & -1.47 & Y & 2, 4, 6, 7\\
GC & \object{NGC~6101}   & \object{Caldwell 107} & 1.0 & 8.00 & 0.13  & 12.5 & -1.76 & Y & 2, 4, 7, 11\\
GC & \object{NGC~6121}   & M4       & 1.17 & 4.7 & 0.25  & 13.1 & -1.98 & Y & 1, 2, 3, 4\\
GC & \object{NGC~6139}   &             & 3.80 & 4.2 & 0.90  & 11.6 & -1.58 & Y & 1, 2, 12 \\
GC & \object{NGC~6144}   &             & 1.69 & 7.17 & 0.24  & 13.8 & -1.56 & Y & 2, 4, 6, 7\\
GC & \object{NGC~6171}   & M107   & 1.17 & 5.5 & 0.21  & 13.4 & -1.03 & Y & 1, 2, 4, 13, 14,\\
GC & \object{NGC~6205}   & M13     & 7.75 & 5.93 & 1.31  & 11.7 & -1.33 & Y & 2, 4, 6, 7\\
GC & \object{NGC~6218}   & M12     & 0.74 & 4.2 & 0.18  & 13.4 & -1.43 & Y & 1, 2, 3, 4\\
GC & \object{NGC~6254}   & M10     & 1.53 & 4.2 & 0.36  & 12.4 & -1.57 & Y & 1, 2, 4, 15\\
GC & \object{NGC~6304}   &             & 2.17 & 4.14 & 0.52  & 13.6 & -0.66 & Y & 2, 4, 6, 7\\
GC & \object{NGC~6341}   & M92     & 4.89 & 4.19 & 1.17  & 13.2 & -2.16 & Y & 2, 4, 6, 7\\
GC & \object{NGC~6352}   &             & 0.37 & 5.68 & 0.065  & 12.7 & -0.70 & Y & 2, 4, 7, 16\\
GC & \object{NGC~6362}   &             & 0.81 & 7.70 & 0.11  & 13.6 & -0.99 & Y & 2, 4, 7, 16\\
GC & \object{NGC~6366}   &             & 0.30 & 5.05 & 0.058  & 13.3 & -0.73 & Y & 2, 4, 7, 16\\
GC & \object{NGC~6388}   &             & 12    & 2.5 & 4.8    & 11.7 & -0.45 & Y & 1, 2, 4, 17\\
GC & \object{NGC~6397}   &             & 1.1   & 3.3 & 0.33  & 13.4 & -1.99 & Y & 1, 2, 4, 18\\
GC & \object{NGC~6441}   &             & 9.55 & 3.3 & 2.92  & 11.2 & -0.44 & Y & 1, 2, 3, 4\\
GC & \object{NGC~6496}   &             & 2.00 & 5.70 & 0.35  & 12.4 & -0.70 & Y & 2, 4, 6, 7\\
GC & \object{NGC~6535}   &             & 0.20 & 2.86 & 0.070  & 10.5 & -1.51 & Y & 2, 4, 6, 7\\
GC & \object{NGC~6541}   &             & 5.72 & 3.93 & 1.45  & 12.9 & -1.53 & Y & 2, 4, 6, 7\\
GC & \object{NGC~6584}   &             & 3.03 & 4.87 & 0.62  & 11.3 & -1.30 & Y & 2, 4, 6, 7\\
GC & \object{NGC~6624}   &             & 2.57 & 3.20 & 0.80  & 12.5 & -0.70 & Y & 2, 4, 6, 7\\
GC & \object{NGC~6637}   & M69     & 2.00 & 3.66 & 0.55  & 13.1 & -0.78 & Y & 2, 4, 7, 19\\
GC & \object{NGC~6652}   &             & 1.09 & 2.37 & 0.46  & 12.0 & -0.97 & Y & 2, 4, 6, 7\\
GC & \object{NGC~6656}   & M22     & 6.44 & 5.32 & 1.21  & 12.7 & -1.49 & Y & 2, 4, 6, 7\\
GC & \object{NGC~6681}   & M70     & 1.79 & 3.16 & 0.57  & 12.8 & -1.35 & Y & 2, 4, 6, 7\\
GC & \object{NGC~6715}   & M54     & 12.9 & 10.7 & 1.20  & 10.8 & -1.25 & Y & 2, 3, 4, 7\\
GC & \object{NGC~6717}   &             & 0.48 & 2.39 & 0.20  & 13.2 & -1.09 & Y & 2, 4, 6, 7\\
GC & \object{NGC~6723}   &             & 3.57 & 6.58 & 0.54  & 13.1 & -0.96 & Y & 2, 4, 6, 7\\
GC & \object{NGC~6752}   &             & 2.82 & 3.8 & 0.75  & 13.8 & -1.55 & Y & 1, 2, 3, 4\\
GC & \object{NGC~6779}   & M56     & 2.30 & 5.11 & 0.45  & 13.7 & -2.00 & Y & 2, 4, 6, 7\\
GC & \object{NGC~6809}   & M55     & 0.55 & 7.6 & 0.07   & 13.8 & -1.93 & Y & 1, 2, 3, 4\\
GC & \object{NGC~6838}   & M71     & 0.20 & 3.3 & 0.06   & 12.7 & -0.82 & Y & 1, 2, 3, 4\\
GC & \object{NGC~6934}   &             & 2.95 & 5.32 & 0.55   & 11.1 & -1.32 & Y & 2, 4, 6, 7\\
GC & \object{NGC~6981}   & M72     & 1.68 & 7.82 & 0.21   & 10.9 & -1.21 & Y & 2, 4, 6, 7\\
GC & \object{NGC~7078}   & M15     & 5.13 & 5.1 & 1.00   & 13.6 & -2.33 & Y & 1, 2, 3, 4\\
GC & \object{NGC~7089}   & M2     & 5.75 & 6.03 & 0.95   & 11.8 & -1.31 & Y & 2, 3, 4, 7\\
GC & \object{NGC~7099}   & M30     & 1.45 & 4.1 & 0.35   & 14.6 & -2.33 & Y & 1, 2, 3, 4\\
{\bf GC} & {\bf\object{Ruprecht~106}} &          & {\bf 0.68} & {\bf 11}  & {\bf 0.06}    & {\bf 12}    & {\bf -1.5}  & {\bf N}  & {\bf 2, 20} \\ 
GC & \object{Terzan~7}     &             & 0.39 & 8.7 & 0.04   & 7.4   & -0.12 & N & 1, 2, 21, 22\\
\hline
\end{tabular}
\end{table*}
%%%%%%%%%%%%%%%%%%%%%%%%%%%%%%%%%%%%%%%%%%%%%%%%%%%%%%%%%%%%%%
%%%%%%%%%%%%%%%%%%%%%%%%%%%%%%%%%%%%%%%%%%%%%%%%%%%%%%%%%%%%%%
% Data for star clusters that have been searched for the anticorrelation
%%%%%%%%%%%%%%%%%%%%%%%%%%%%%%%%%%%%%%%%%%%%%%%%%%%%%%%%%%%%%%
\setcounter{table}{3}
\begin{table*}[t]
\caption{continued.}             
\label{t:scdata}      % is used to refer this table in the text
\centering                          % used for centering table
\begin{tabular}{l c c c c c c c c l}        % centered columns (4 columns)
\hline\hline                 % inserts double horizontal lines
\vspace*{.1mm}
Class & Object & Alt. name & 
$M_*$& $r_\mathrm{h}$\tablefootmark{a} & 
$C_5$\tablefootmark{b} & Age & $\sbr{\mathrm{Fe/H}}$ & MPs\tablefootmark{c} & Ref.\\[.2mm]
&&& $(10^5\msm)$ & (pc) &  &(Gyr)& & & \\
\hline
GC & \object{Palomar~12}  &            & 0.28 & 16  & 0.024  & 8.6   & -0.81 & N & 1, 2, 21\\
\\%\hline
OC & \object{Berkley~39}   &             & 0.2   & 11  & 0.018   & 6   & - 0.21 & N  & 23, 24 \\
OC & \object{Collinder~261}&           & 0.01 & 2\tablefootmark{d}&  0.005 &  6  & -0.03  & N  & 24, 25, 26 \\
OC & \object{Melotte~25} &\object{Hyades}& 0.004 & 3.7 & 0.001  & 0.6 &  +0.13 & N  & 27, 28, 29, 30\\
OC & \object{Melotte~111}  & \object{Coma Berenices} & 0.001 &3.5&0.0003&0.5&+0.07&N & 27, 29, 31, 32 \\
OC & \object{NGC~~~752} &             & 0.001 & 1.4     & 0.0009 & 1.12  & -0.02 & N  & 29, 33, 34\\ 
OC & \object{NGC~1817} &  \object{Collinder~4a} & 0.02 & 7     & 0.002 & 1  & -0.11 & N  & 27, 32, 34, 35, 36\\ 
OC & \object{NGC~2360} &  \object{Melotte~64} & 0.018 & 2     & 0.009 & 0.56  & -0.07 & N  & 29, 34, 37\\ 
OC & \object{NGC~2506} &  \object{Collinder~170} & 0.03 & 15 & 0.002 & 1.11  & -0.19 & N  & 29, 34, 38, 39\\
OC & \object{NGC~2682}   & \object{M67}   & 0.01 & 2.5 & 0.004 & 4.4 & +0.05 &  N &  29, 40, 41 \\
OC & \object{NGC~3114} &      & 0.0016 & 8 & 0.0002 & 0.16  & -0.01 & N  & 24, 29, 36, 42 \\
OC & \object{NGC~6134}   &             &  0.0008 & 2 & 0.0004 & 0.7   & + 0.12 & N & 24, 29, 43, 44\\
OC & \object{NGC~6253} & \object{Melotte~156} & 0.007 & 2 & 0.0035 & 4 & +0.4 & N & 24, 26, 29 \\
OC & \object{NGC~6475} & \object{M~7} &0.007  &  6 &  0.001 &  0.3 & +0.14 & N & 29, 32, 36 \\%checked-ok up to here
OC & \object{NGC~6705}   &           & 0.1  &  1.2 &  0.08  & 0.3 & +0.10 & N & 21\\
OC & \object{NGC~6791}   &  \object{Berkeley 46} & 0.5 & 5 & 0.1 & 7.5 & +0.30 & N & 29, 45, 46 \\
OC & \object{NGC~7789} & \object{Melotte~245} & 0.06 & 5 & 0.01 &1.6 & +0.04 & N & 29, 36, 41, 47\\
OC & \object{IC~4651}    & \object{Melotte~169} & 0.006& 2 & 0.003& 1.1& +0.15& N & 24, 29, 36, 48 \\
\\%hline
Fornax & \object{For~3}     &             & 5.2  & 7.2  & 0.72    &  13     &   -2.33 & Y  &  49\\
Fornax & \object{For~5}     &             & 3.2  & 8.5  & 0.38    &  13    &   -2.09  & Y  &  49\\
\\
LMC & \object{NGC~1651} &            & 0.8  & 22 & 0.04 & 1.5 & -0.3 & N & 50, 51, 52 \\
LMC & \object{NGC~1783} &            & 2.5  & 19 & 0.13 & 1.4 & -0.3 & N & 50, 51, 52 \\
LMC & \object{NGC~1786} &            & 3.7 & 6.0 & 0.62 & 13 & -1.75 & Y & 53, 54, 55 \\ %rh claculated in emf_rh.pro
LMC & \object{NGC~1806} &            & 1.3  & 15 & 0.08 & 1.4 & -0.3 & N & 50, 56, 57 \\
LMC & \object{NGC~1846} &            & 1.7  & 15 & 0.11 & 1.4 & -0.3 & N & 50, 56, 58 \\
LMC & \object{NGC~1866} &            & 0.8  & 19 & 0.04 & 0.2 & -0.3 & N & 50, 59, 60, 61 \\
LMC & \object{NGC~1978} &            & 2.0  & 8.7   & 0.23 & 2.0 & -0.3 & N & 50, 51, 62, 63, 64, 65\\
LMC & \object{NGC~2173} &            & 0.5  & 11 & 0.05 & 1.6 & -0.3 & N & 50, 51, 52 \\
LMC & \object{NGC~2210} &            & 3.0 & 5.9 & 0.51 & 13 & -1.65 & Y & 53, 54, 55 \\%rh claculated in emf_rh.pro
LMC & \object{NGC~2257} &            & 2.6 & 19 & 0.14 & 13 & -1.95 & Y & 53, 54, 55 \\%rh claculated in emf_rh.pro
\\
\multicolumn{3}{l}{Scutum red supergiant cluster~1}  & 0.32 & 2.6 & 0.12   & 0.012 & -0.15 & N & 66, 67 \\
\multicolumn{3}{l}{Scutum red supergiant cluster~2}  & 0.40 & 4.6 & 0.087 & 0.017 & -0.15 & N & 66, 67 \\
\hline
\end{tabular}
\tablefoot{
\tablefoottext{a}{Half-mass radii were computed from the half-light radii \rv{(apparent size was taken as equivalent to half-light radii)}
	from the literature, assuming a Plummer model for the cluster. }
\tablefoottext{b}{Compactness $M_*/r_\mathrm{h}$ in units of $10^5\msm/$pc.}
\tablefoottext{c}{Y: possesses \rv{multiple populations}, N: \rv{multiple populations have} 
	been searched for, but not found.}
\tablefoottext{d}{Estimated from Fig.~5 in ref. 16.}
}
\tablebib{
	(1)~\citet{Carea10a}; 
	(2)~\citet[2010 edition]{Harris96};  
	(3)~\citet{Kimmea15}; 
	(4)~\citet{Piottoea15}:
	(5)~\citet{LebzW11};
	(6)~\citet{Boylea11};
	(7)~\citet{ForBri10};
	(8)~\citet{Kolea08};
	(9)~\citet{Luetzea13}; 
	(10)~\citet{Mucciea15a};
	(11)~\citet{CalDAnt11};
	(12)~\citet{Bragaglea15a}
	(13)~\citet{Conroy12}; 
	(14)~\citet{Piatea94}; 
	(15)~\citet{ZBV12}
	(16)~\citet{Mandushea91};
	(17)~\citet{Lanzea13};
	(18)~\citet{Heylea12}; 
	(19)~\citet{Miocchi07}
	(20)~\citet{Villanea13a};
	(21)~\citet{Cantea14}; 
	(22)~\citet{SalWei02};
	(23)~\citet{Bragaglea12a}; 
	(24)~\citet{Tadross01}
	(25)~\citet{BragagliaT06}
	(26)~\citet{Mikolaitea12};
	(27)~\citet{Casewea2014};
	(28)~\citet{Lodiea14};
	(29)~\citet{MacLeanea15}; 
	(30)~\citet{PZMcMH01}
	(31)~\citet{GMR08}
	(32)~\citet{BF71};
	(33)~\citet{Francic89}; 
	(34)~\citet{Reddea12};
	(35)~\citet{Donatea14};
	(36)~\citet{Gozea12};
	(37)~\citet{GKB12};
	(38)~\citet{LKA13}; 
	(39)~\citet{MarconiGea97};
	(40)~\citet{Hurlea05};
	(41)~\citet{Pancinea10}; 
	(42)~\citet{SPD13}; 
	(43)~\citet{Smiljea09}
	(44)~\citet{Mikolaitea10}
	(45)~\citet{Bragaglea14a} 
	(46)~\citet{Platea11};
	(47)~\citet{Wu09}
	(48)~\citet{Mikolaitea11};
	(49)~\citet{LBS12};
	(50)~\citet{Goudfrea14};
	(51)~\citet{Mucciea08a}; 
	(52)~Niederhofer et al. in prep;
	(53)~\citet{MackGil03}	
	(54)~\citet{Mucciea09a}; 
	(55)~\citet{Mucciea10a}	
	(56)~\citet{BN15}; 
	(57)~\citet{Mucciea14a}; 
	(58)~Mackey et al. (in prep.); 
	(59)~\citet{BasSil13};
	(60)~\citet{Mucciea11a}; 
	(61)~\citet{Niedea15}; 
	(62)~\citet{Ferraroea06};
	(63)~\citet{Milonea09}; 
	(64)~\citet{Mucciea07a}; 
	(65)~\citet{WerchZar11};
	(66)~\citet{DaviesBea09};
	(67)~\citet{PZMcMG10};

%	(67)~\citet{Balbinea09} NGC6642
%}
}
\end{table*}
%%%%%%%%%%%%%%%%%%%%%%%%%%%%%%%%%%%%%%%%%%%%%%%%%%%%%%%%%%%%%%

%%%%%%%%%%%%%%%%%%%%%%%%%%%%%%%%%%%%%%%%%%%%%%%
\subsection{Gas expulsion and multiple populations}
%%%%%%%%%%%%%%%%%%%%%%%%%%%%%%%%%%%%%%%%%%%%%%%

\subsubsection{Constraints from N-body simulations}
\rv{\citet{KhalBaum15} show N-body simulations of star clusters that contain 
chemically normal (first generation) and peculiar (second generation) stars.
With a simple model of gas expulsion they find that star formation efficiencies 
$\lesssim 50$~per cent are needed in order to produce star clusters with final ratios
of first to second generation stars close to the ones observed in present-day
globular clusters in the Milky Way. Our model grid results imply that
this would only be possible if the  compactness indices $C_5$ would not exceed
unity (or up to 30, if series of strong hypernovae would occur in such clusters).
}

\rv{We collected parameters for clusters that have been searched 
for \rv{multiple populations} from the literature (Table~\ref{t:scdata}),
and calculated their $C_5$ value.
We included both, clusters with detailed abundance analysis, as well as ones that show
multiple populations in the colour-magnitude diagrams.}

\rv{Current $C_5$-values extend up to a few for objects that do show 
\rv{multiple populations (they are all} globular clusters). 
This would already put them into a regime where
gas expulsion by stellar winds and supernovae would not have been possible. However,
their masses must have been even higher, and the radii smaller, if gas expulsion
would indeed have been at work in these objects:
\citet{BL15} compare the first to second generation ratios predicted
by \citet{KhalBaum15} to observations and find that all clusters that show
the Na-O anticorrelation should have lost a very similar fraction of stars,
between 95 and 98 per cent of the initial number
\citep[compare also][]{PC06,Carea10a,DErcolea10}. Expansion factors
are up to a few \citep{KhalBaum15}. The current $C_5$-values,
0.05-3.2, would hence transform to the range of about 1 to 100
for the initial compactness indices.
 Hence, even a series of very
energetic events, like hypernovae, could not 
have been the agent of gas expulsion in the most compact objects. 
This is in the context of early gas expulsion 
models, motivated by the compact, young, gas-free clusters discussed earlier.
Energetic events late in the evolution of a star cluster
have been discussed elsewhere
\citepalias{Krausea12a}, and would also be a possible agent for
gas expulsion in young GCs, if the results from YMCs (gas-free after a few Myr)
would for some reason not apply to clusters that show light-element 
anticorrelations.}

\rv{Summarising, these results suggest that gas expulsion is not a viable solution to the
mass budget problem in the context of multiple population star clusters. A similar 
conclusion was reach with different arguments by \citep{KhalBaum15} and \citet{BL15}.
}
%%%%%%%%%%%%%%%%%%%%%%%%%%%%%%%%%%%%%%%%%%%%%%%%%%%%%%%%%%%%%%
% Critical SFE compared to observations: multiple populations in mass-age-met-C5 
%%%%%%%%%%%%%%%%%%%%%%%%%%%%%%%%%%%%%%%%%%%%%%%%%%%%%%%%%%%%%%
\begin{figure*}
  \centering
	%\sidecaption  
\includegraphics[width=0.495\textwidth]{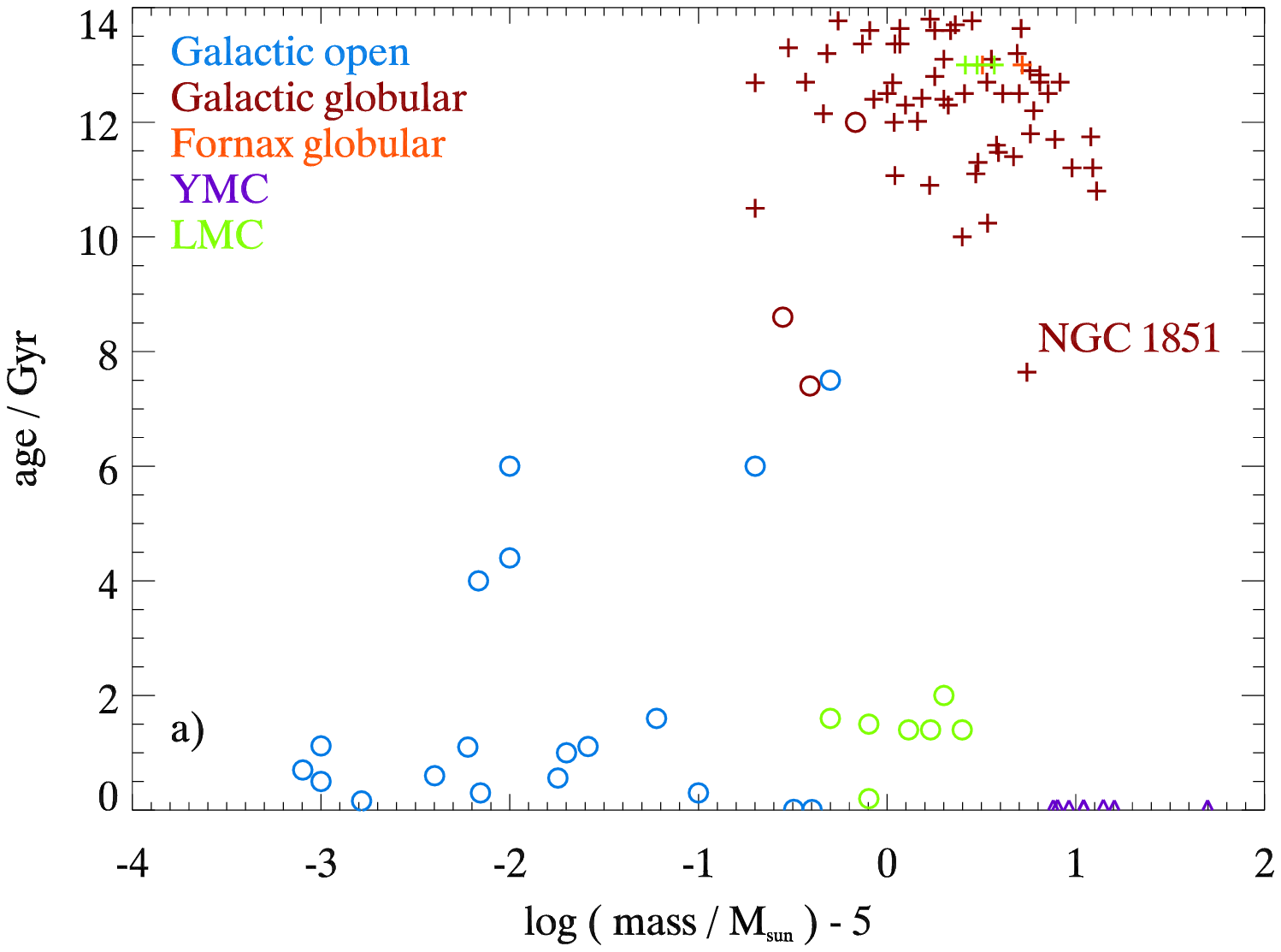}  
\includegraphics[width=0.495\textwidth]{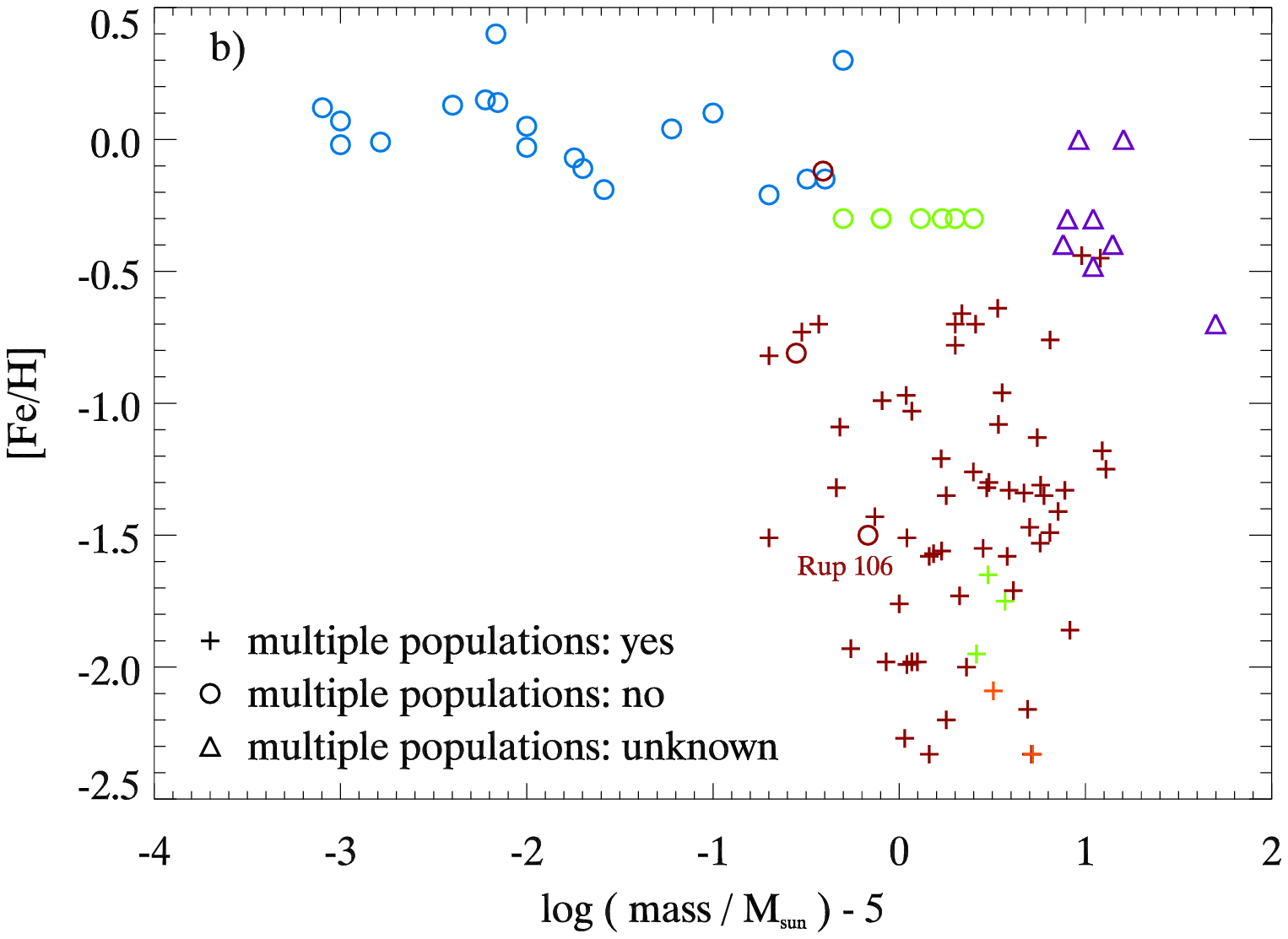}  
\includegraphics[width=0.495\textwidth]{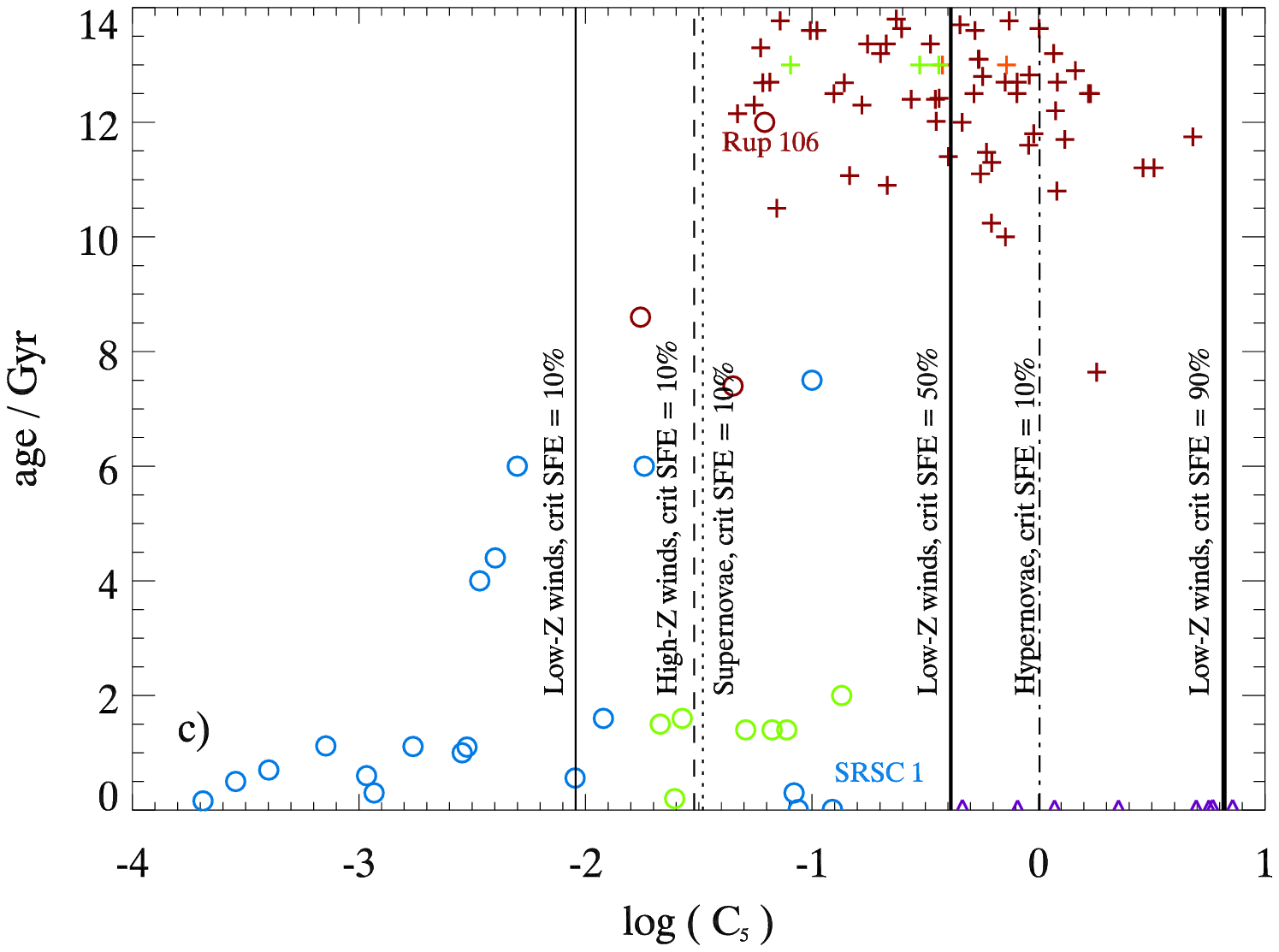}  
\includegraphics[width=0.495\textwidth]{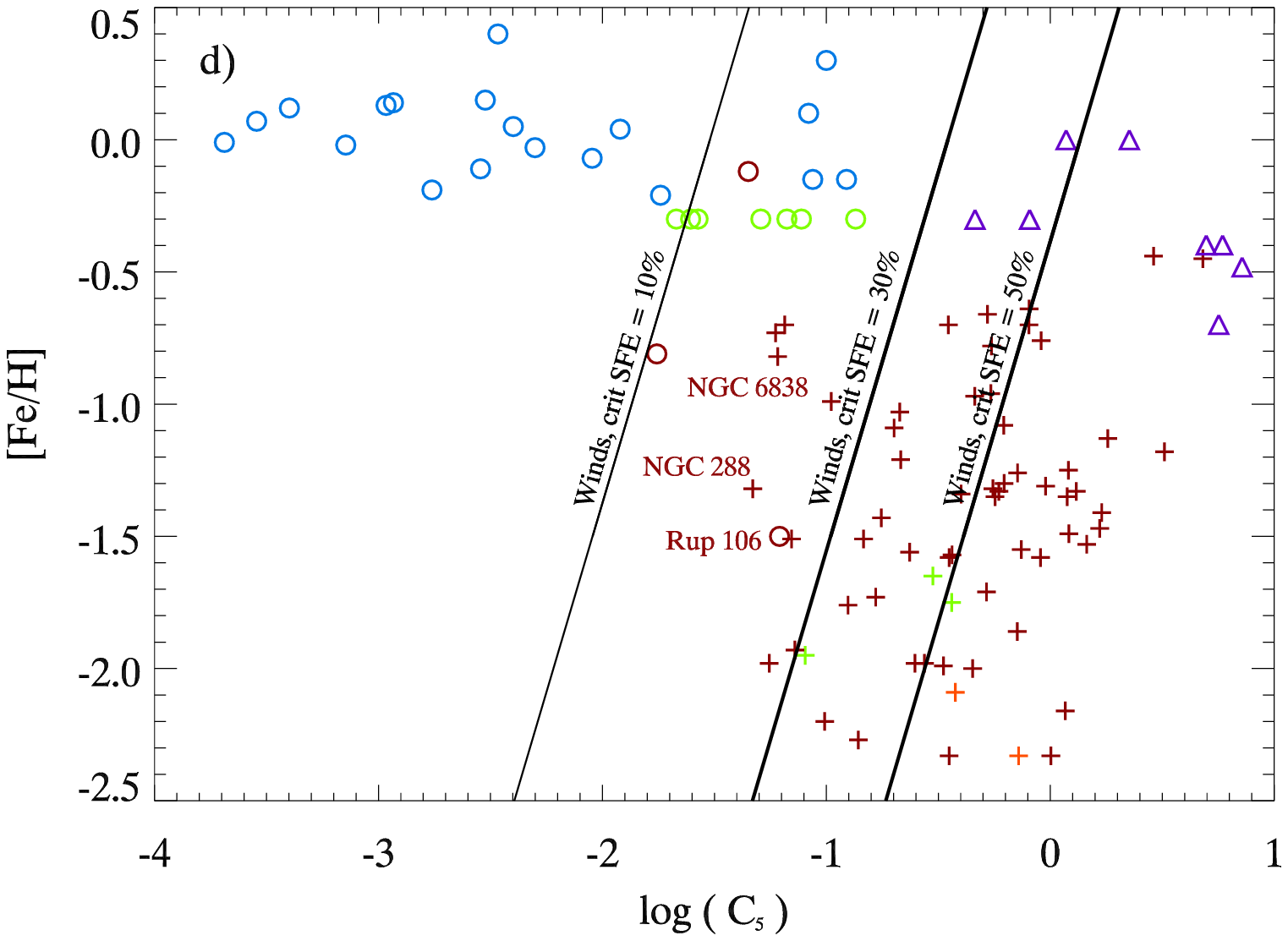}  
  \caption{\rv{S}tar clusters 
	from Tables~\ref{t:YMCs} and~\ref{t:scdata} in \rv{different parameter spaces}.
	Different symbols are used according to the presence of \rv{multiple populations, 
	as indicated either from photometry or spectroscopic abundance analysis}:
	open circles, plus-signs and triangles indicate, respectively, star clusters that have 
	been searched for \rv{multiple populations but with negative result}, ones 
	\rv{with positive result}, and ones where the presence of \rv{multiple populations} 
	is not known. The 
	colour\rv{s} code different samples: GCs: dark red, Fornax globular clusters: light red,
	YMCs: purple, LMC clusters: green, Galactic open clusters: blue. 
	\rv{Some star clusters discussed in the text are highlighted. 
	a) Mass-age diagram. b) Mass-metallicity diagram. c) Compactness-age diagram.}
 	Black vertical lines
	indicate the critical star formation efficiency required
	in order to expel the remaining gas 
	as calculated in the present work. 
	Solid lines are for the low metallicity stellar winds 
	models (thin: 10 \rv{per cent}, medium: 50 \rv{per cent}, thick: 90 \rv{per cent}). 
	\rv{O}ther lines are for 10~\rv{per cent} critical star formation efficiency;
	the dashed one for solar metallicity winds, the dotted one for normal supernovae,
	and the dot-dashed one for hypernovae of $10^{53}$ erg, each. 
	\rv{d) Compactness-metallicity diagram. Solid lines separate regions where gas 
	expulsion is possible with
	the indicated star formation efficiency (left) from regions where this is not possible 
	(right), according to the present work, for the case of metallicity dependent stellar winds.}
	}
   \label{f:c5-age}%
\end{figure*}
%%%%%%%%%%%%%%%%%%%%%%%%%%%%%%%%%%%%%%%%%%%%%%%%%%%%%%%%%%%%%%

\subsubsection{Parameter space for abundance anomalies}
\rv{We compare the parameter space occupied by star clusters with and without evidence for
multiple populations to the ones where gas expulsion may occur in more detail in Fig.~\ref{f:c5-age}.
Apart from a possible early gas expulsion event, s}tar clusters lose 
mass over time due to, e.g., low-mass star depletion \citep{KM09},
encounters with the environment \citep{Kruijea12c}, or disc passages \citep{Webbea14},
which may also lead to expansion via tidal heating. Hence, we generally expect that
$C_5$ will decrease with time,  
\rvtwo{easily by a factor of a few and very likely}
in different ways for different objects 
\rvtwo{\citep[e.g.,][]{RosHur15a}. For example,
\citet{RosHur15b} reconstructed the mass-loss history of the two Milky Way GCs
\object{HP~1} and \object{NGC~6553}. According to these calculations,
the former GC experienced a secular mass loss of 79~per~cent of its initial mass, the latter
one even 93~per~cent.}
\rvtwo{Recent observational studies typically have abundance uncertainties less
than 0.1~dex \citep[e.g.,][]{Mucciea11a,Villanea13a}.
Mass, age and radius accuracies are of the order of 10~per~cent \citep{Bastea14b}.
Absolute ages for GCs have a systematic uncertainty \citep{Carea10a}.}
\rvtwo{Therefore, systematic effects should dominate, but measurement uncertainties
might also be important in some cases, especially for the older studies.}

\rv{The mass-age diagram is shown in Fig.~\ref{f:c5-age}a. With the addition of the 
relatively recent data from the 
Large Magellanic Cloud (LMC, green), the overlap in mass between clusters that do or do 
not have multiple populations has become substantial. It is now more than one dex, 
suggesting that at least a second parameter is required. The latter could be age, with multiple populations
being restricted to old star clusters. Clear outliers would be the single population cluster 
\object{Rup~106} with an age of 12~Gyr, and the multiple population cluster \object{NGC~1851} at 8~Gyr.}

\rv{The second parameter could also be metallicity (Fig.~\ref{f:c5-age}b). In this case,
star clusters with $\sbr{\mathrm{Fe/H}}$ below -0.5 would have multiple populations. Outliers in this case would be 
the single population clusters \object{Pal~12} (\feh{-0.8}) and \object{Rup~106} (\feh{-1.5}). Especially
the latter is located far inside the region of parameter space occupied by multiple-population clusters.
}

\rv{When considering the compactness index $C_5$ instead of the mass as 
fundamental parameter for the expression
of multiple populations (Fig.~\ref{f:c5-age}c), star clusters with and without 
multiple populations are also not cleanly separated. The overlap is, however,
reduced: The most compact single population cluster in our sample is the Scutum red 
supergiant cluster with  log($C_5$)=$-0.92$. The least compact
multiple population cluster is \object{NGC~288} with log($C_5$)=$-1.30$. 
Thus, the overlap region is much smaller compared
to the case where mass is taken as the primary parameter. Again, age may 
be taken as additional parameter to separate 
single and multiple population clusters with the same caveat as above.}

\rv{Star clusters with and without multiple populations separate more clearly in the 
compactness-metallicity diagram (Fig.~\ref{f:c5-age}d) compared to the mass-metallicity diagram.
In particular, \object{Rup~106} has now moved towards the left border of the region occupied
by multiple population clusters.}

\rvtwo{
\object{Rup~106} is a particularly well studied object \citep{Dottea11,Villanea13a}:
Its age is accurately known from Hubble Space Telescope 
ACS\footnote{Advanced Camera for Surveys}
photometry. Abundances for nine stars have been measured with the high resolution 
spectrograph UVES\footnote{Ultraviolet and Visual Echelle Spectrograph} 
on the Very Large Telescope. The probability to have missed a second population
with similar properties than in other GCs is below $10^{-4}$.
Hence, it is not an outlier with large observational uncertainties.
}

\rv{Critical star formation efficiencies for gas expulsion are indicated in Fig.~\ref{f:c5-age}c and d.
Considering current parameters, single population clusters prefer regions of parameter 
space where gas expulsion would be
easily possible from our analysis. Multiple population star clusters extend into regions
of parameter space, where stellar wind and supernova feedback could only remove 
the residual gas when
the star formation efficiency was above 50 per cent -- too much for gas expulsion with 
significant loss of stars. This conclusion is reinforced when considering that evolutionary effects
tend to increase the radii and decrease the masses of star clusters, 
i.e. they have typically been more compact in the past. }

\rv{The critical star formation efficiencies for metallicity dependent stellar winds is 
indicated by the oblique lines in Fig.~\ref{f:c5-age}d. These lines again do not cleanly
separate single from multiple population clusters. They do, however, further
reduce the overlap in $C_5$ from 0.38 dex to about 0.3 dex.
Hence, a scenario, where multiple populations are suppressed whenever the compactness
is low enough so that stellar winds can expel the gas, best explains the data. 
%A clear prediction is that the YMCs in our
%sample should all have multiple populations.
}
 
\rvtwo{
It is instructive to compare \object{Rup~106} to \object{NGC~6535}. 
\object{Rup~106} has been mentioned above as a well-studied single population GC.
It is marked in Fig.~\ref{f:c5-age}d, and sits marginally, but clearly, in the parameter space occupied by multiple-population clusters. \object{NGC~6535} is a multiple-population
cluster with essentially the same $C_5$ and metallicity than \object{Rup~106}.
Consequently, their symbols partially overlap in Fig.~\ref{f:c5-age}d.
\object{NGC~6535} has a Galactocentric distance \citep{Harris96} of 3.9~kpc, hence
can be expected to have experience strong mass loss due to tidal interactions with disc and 
bar \citep{RosHur15b,RosHur15a}. Its initial $C_5$ was therefore likely much higher, i.e.
it would have been found more to the right, in the cloud of the other multiple-population
clusters. \object{Rup~106} has a Galactocentric distance of 18.5~kpc. Therefore,
secular mass loss was probably lower, and its initial location on Fig.~\ref{f:c5-age}d
was likely not very different from the current one.
The situation is similar for the multiple-population GCs with similar $C_5$ and higher
metallicity than these two objects (pluses above them in Fig.~\ref{f:c5-age}d):
They are all at rather low Galactocentric distances (3.3 to 6.7) kpc. An exception is
\object{NGC 288} at 12~kpc, also accurately measured.
}

\rvtwo{These considerations are consistent with the aforementioned expectation that
systematic errors due to secular evolution dominate the uncertainties. The small overlap of
0.3 dex found above for a separation due to gas expulsion by stellar wind feedback agrees
quantitatively with the expected magnitude of the systematic uncertainties.
Hence, the hypothesis describes the data well within the general limitations mentioned above. }

The YMCs from the sample of \citet{Bastea14b} would form a powerful test for 
\rv{a possible connection between gas expulsion and the expression of
multiple populations.}
Some of these objects have even higher $C_5$ than the most compact
GCs. They also join the GCs smoothly in the $\sbr{\mathrm{Fe/H}}$ over $C_5$ plot, so that one would expect them
to show anticorrelations, based on these parameters. If they would not show the anticorrelation, it would be clear
that the physics of the clearing of the residual gas is not related to it
or that YMCs can not be considered as the counterpart of young GCs. 

%%%%%%%%%%%%%%%%%%%%%%%%%%%%%%%%%%%%%%%%%%%%%%%
\subsection{Accuracy of the method}\label{s:dacc}
%%%%%%%%%%%%%%%%%%%%%%%%%%%%%%%%%%%%%%%%%%%%%%%
We presented a number of thin shell models for gas expulsion, where
gas expulsion was deemed successful, if the solution passed a 
Rayleigh-Taylor criterion. 

These models are simplifications of the
3D~hydrodynamic problem that allowed the calculation of a large
number of models. A 3D initial condition will entail some degree of 
clumping of the gas. During the early embedded phase, 
such gas is likely to continue gravitational contraction,
whereas the diffuse component is shifted into shells, as also 
seen in 3D hydrodynamics 
simulations of low-mass clusters \citep[e.g.,][]{DB12}. 
The clumped gas that is about to form stars 
contributes to the gravitational potential without producing feedback energy,
which makes it more difficult for the remaining gas to escape. On the other hand,
there is effectively less gas to expel, which makes it easier for the given amount
of feedback energy to accelerate the reduced gas mass.  
The overall effect might thus not be very strong. However,
3D simulations are now necessary to assess this quantitatively. 
If the gas would remain in the clusters for long enough to be affected 
by supernovae (few Myr), the denser clumps would likely have formed
stars or protostellar cores already, and the remaining gas should probably
be reasonably smooth, such that spherical symmetry would again likely be
a reasonable assumption. 

The biggest uncertainty probably concerns the
initial conditions. The assumption that the gas is distributed radially
like the stars is not likely to hold. For hydrostatic equilibrium
it would require pressure support. Thermal pressure would be unlikely due
to strong cooling expected at the implied gas densities. Magnetic fields
would probably lead to compression into filaments, and then to star formation.
Radiation pressure has been shown in Sect.~\ref{s:method} to be insufficient.
Turbulence could support the gas for about one crossing time \citep[e.g.,][]{ES04}.
Hence, we expect the gas generally to be in an inflow or outflow state.

During the formation of the stars and shortly afterwards, we expect the 
gas-to-star ratio in the centre of the cluster to be somewhat lower than what we
assumed, due to gas depletion by star formation \citep[e.g.,][]{Kruijea12a}. 
If the gas is not expelled 
early, then we might expect the 
gas-to-star ratio in the centre to be higher than what we assumed later on 
due to accumulation in the centre.
Generally, the gas is likely to have an inward velocity, because it is, of course,
accumulating to form a star cluster.
Except perhaps for the earliest phases, these effects would make gas expulsion 
more difficult than implied by the present models.

\rv{Some recent 3D hydrodynamics simulations have addressed already a few issues present in our simple models:
Distributed energy / momentum sources lead to enhanced dissipation
\citep{Krausea13a,BP16}. This is again a factor that makes it easier
for gas expulsion to occur in our simple models, where energy input effectively
happens at the cluster centre.}

\rv{\citet{Calurea15} simulate an embedded star cluster with compactness index $C_5 = 0.85$
and a metallicity of \feh{-1.3}.
For these parameters, our simple model predicts a minimum local star formation efficiency
for gas expulsion on the crossing timescale of 0.61 in the wind phase, and, respectively 
0.41 in the supernova phase. The authors use a star formation efficiency of 0.4.
They do not find gas expulsion on the crossing timescale, but a slow mass loss with a characteristic
timescale of 10~Myr, entirely consistent with the prediction of our simple model.}

We are thus confident that our models are able to reasonably rule out gas expulsion for
certain star cluster parameters. However, star clusters for which our model 
predicts gas expulsion might still be able to retain the gas, for example, 
because of a more concentrated gas distribution than assumed in the model.

\section{Conclusions}\label{s:conc}

\rv{The gas expulsion paradigm for massive star clusters is currently 
debated in the literature.}
We investigated \rv{the conditions for} 
gas expulsion in massive star clusters in the context of 
different observational constraints by means of a large number
of self-gravitating thin-shell models and an analytical estimate for the 
stability of the shell.

\rv{For several extragalactic, young massive clusters, we found that gas expulsion would 
not have been possible for standard assumptions, including an
\rvtwo{IMF} according to \citet{Kroupea13},
a star formation efficiency of 30~per cent, and the assumption that 
20~per cent of the energy produced by stellar feedback couples to the gas.
The result holds for both, stellar winds and 
supernovae as energy sources. Radiation pressure is not important in this context.
Yet, these clusters are gas-free at an age of
around 10~Myr.}

\rv{We then showed that the gas could be cleared if the star formation efficiency
would be increased. For some clusters the star formation efficiencies
would have to be higher than 80~per cent to remove the rest of the gas.
This, however, would be too high for any significant change of the gravitational
potential. Hence, the stellar mass could not be significantly affected
by this kind of gas loss. Thus, for canonical energy production and
\rvtwo{IMF}, the standard paradigm of residual gas expulsion
(implying a substantial loss of stars)
cannot work in these clusters.
 }

%\rv{We then investigated reasonable modifications of the energy production: 
%hypernovae, related to the formation of compact objects might
%be able to still lead to gas expulsion with star formation efficiencies around 
%30~per cent. 
%%The requirement that not just one, but many hypernovae would
%%be required to expel the gas in a given cluster, makes this scenario perhaps unlikely.
%}

\rv{With a grid of star clusters with masses between $10^5\msm$
and $10^7\msm$, and half-mass radii between 1 and 10 pc,
we showed that gas expulsion, if successful, indeed happens on the
crossing timescale. This confirms that the supershell ansatz is indeed
a useful realisation for gas expulsion on the crossing timescale.}

\rv{T}he critical star formation efficiency 
\rv{required} to remove the \rv{remaining} gas 
is a function of the compactness index $C_5\equiv  (M_*/10^5 M_\odot)/(r_\mathrm{h}/\mathrm{pc})$. 
\rv{More compact clusters (higher $C_5$) would transform
relatively more gas into stars. Any gas loss in massive, compact star clusters would
therefore be moderate, and could not lead to strong losses of stars, in good agreement
with a recent census of GCs and stars in Fornax \citep{Larsenea14b}.
This is, however, a problem for models that try to explain the observed ratio of first to second generation stars
in globular clusters by a gas-expulsion induced loss of first generation stars.
}

\rv{We showed by comparison to observations that compactness is a better predictor
for the occurrence of multiple populations than mass. This suggests that the absence
of gas expulsion might play a role for the expression of the Na-O anticorrelations
and multiple populations in globular clusters. The best separation between
clusters that do and ones that do not have multiple populations is achieved by 
a model for gas expulsion based on metallicity dependent stellar winds.}

\rv{Gas expulsion scenarios that have been developed to reproduce the light
element anticorrelations in globular clusters would require initial compactness
indices of $1 \lesssim C_5 \lesssim 100$. 
In \citetalias{Krausea12a}, energetic events at
late times were proposed as a solution to this problem.
Observations of YMCs suggest, however, that the gas is lost early in the evolution of
also massive star clusters.
Here, we showed that early
gas expulsion, cannot work for most of this parameter range.
On the basis of this argument, gas expulsion cannot be the solution of the mass budget problem for
the expression of multiple populations in globular clusters.}

\begin{acknowledgements}
  %\rv{We thank the anonymous referee for useful comments.}
We thank Cyril Georgy for providing data for rotating star models,
and Georges Meynet and Andrii Neronov for helpful discussions.
This work was supported by funding from Deutsche Forschungsgemeinschaft
under DFG project number  PR 569/10-1  in the context of the
Priority Program 1573  “Physics of the Interstellar Medium” and 
 the International Space Science Institute (International Team 271 “Massive Star Cluster across the Hubble Time”).
Additional support came from funds from the Munich Cluster of Excellence
“Origin and Structure of the Universe”. (www.universe-cluster.de). 
C.C
acknowledges
financial support from the Swiss National Science Foundation (FNS) for the
project 200020-140346 "New perspectives on the chemical and dynamical evolution
of globular clusters in light of new generation stellar models" (PI C.C.).
We thank the International Space Science Institute (ISSI, Bern, CH) for welcoming
the activities of ISSI Team 271 “Massive Star Clusters across the Hubble
Time" (2013 - 2015; team leader C.C.).
We thank the anonymous referees for very useful comments that helped
to improve the presentation of these results.
%We acknowledge the use of NASA's {\it SkyView} facility
 %    (http://skyview.gsfc.nasa.gov) located at NASA Goddard
 %    Space Flight Center.
\end{acknowledgements}

%.
%
%______________________________________________________________
%
%   \begin{figure} % data level 4
%   \resizebox{\hsize}{!}
%            {\includegraphics{}}
%      \caption{. 
%}
%         \label{fig:fig2new}
%   \end{figure}
%
%______________________________________________________________
%
\bibliographystyle{aa}
\bibliography{/Users/mkrause/texinput/references}

\appendix
\section{Gas pressure versus radiation pressure}\label{ap:rp_vs_thermal}
In our models we consider the effect of gas pressure, only, because the 
radiation pressure is generally too small to lead to gas expulsion in massive clusters.
We verified this by calculating the total luminosity of massive stars ($>8$~\msun)
for the low metallicity \feh{-1.5} FRMS grid of \citet{Decrea07a} as well as 
the model grids of moderately rotating stars at solar metallicity and \feh{-0.8}
from \citet{Ekstrea12a} and \citet{Georgyea13a}, respectively.
For example, a $M_*=10^6$~\msun young star cluster produces a total luminosity
of $L=1\times10^{43}$~erg~s$^{-1}$ during the first few Myr of its lifetime, 
varying by a factor of two for the different model grids. 
For efficient scattering or absorption of the entire spectrum, 
the force exerted by this radiation on a shell outside the distribution of stars
would be $f_1f_2L/c$, where the factor $f_1$ is less than unity and accounts for the 
fact that a part of the stellar distribution is outside the shell, and $f_2$ may exceed
unity and accounts for multiple scatterings.
The energy  transfered to the shell
after the latter would have been lifted to the half-mass radius $r_\mathrm{h}$
would then be 
$E_\mathrm{R}=F_1 F_2 L r_\mathrm{h}/c = 3\times 10^{51} F_1 F_2 r_\mathrm{h,3}$~erg,
where $F_1$ and $F_2$ now refer to path averages of the respective factors
$f_1$ and $f_2$.
This compares to the binding energy of the gas \citep{BKP08},
$E_\mathrm{G}=0.4 (1-\epsilon_\mathrm{SF}) (M_*/\epsilon_\mathrm{SF})^2/r_\mathrm{h}= 
9\times 10^{52} r_\mathrm{h,3}^{-1}$,
where we have used $\epsilon_\mathrm{SF}=0.3$ as an example.
Numerical simulations of radiation pressure driven outflows find efficiency 
factors $f_2$ not significantly above unity \citep{KT13}, due to instabilities
which tend to create optically thin regions where the radiation escapes easily. 
Therefore, the energy delivered by the radiation force does not reach the 
binding energy for this example.
Because the binding
energy scales with $M_*^2$, whereas the energy produced by all
stellar feedback processes including radiation pressure is linear in the mass,
this statement is even stronger for higher masses.

By comparison, using even the low metallicity (\feh{-1.5}) wind powers from the
stellar models of \citet{Decrea07a}, the stellar winds need about 4.5 Myr to 
produce an energy equivalent to the binding energy for the above example. 
Taking into account an efficiency factor of 0.1 (compare Sect.~\ref{s:m:imp}),
they still deliver $1.4\times 10^{52}$~erg within the first 3.5~Myr, i.e. before
the first supernova.

We note that the presence of hot gas in superbubbles has been disputed
recently \citep{MQT10}, based on low X-ray luminosity in some objects
and a possible overprediction of the total X-ray luminosity of star-forming
galaxies. However, 3D hydrodynamics simulations with time-dependent 
energy input have shown that, thanks to 3D instabilities, the X-ray luminosity 
is strongly variable on
Myr timescales and that the integrated X-ray luminosity for star-forming
galaxies is well below the observed values, leaving room for contributions
from other X-ray sources \citep{Krausea14a}. 

Thus, radiation pressure may safely be neglected for the studies in this paper.
It might also be thought of as being included in the stellar wind power
with our adopted efficiency factor of 0.2.  

\end{document}